\documentclass[12pt]{article}

\font\bfcal=eusb10 at 12pt

\newcommand{\be}{\begin{equation}}
\newcommand{\ee}{\end{equation}}
\newcommand{\beqs}{\begin{eqnarray}}
\newcommand{\eeqs}{\end{eqnarray}}

\def\({\left(}
\def\){\right)}

\def\na{\nabla}
\def\da{\dot{\alpha}}
\def\db{\dot{\beta}}
\def\U{\Upsilon}

\def\L{ {\cal L}}
\def\DD{ {\hbox{\bfcal D}}}
\def\Dl{ {\cal D}}

\def\Z{ {\cal Z}}
\def\Q{ {\cal Q}}
\def\Quat{{\bf Q}}
\def\vv{{\bf V}}
\def\WW{{\bf W}}

\def\Tr{{\rm Tr}}
\def\mxth{\mathsurround=0pt }
\def\xversim#1#2{\lower2.pt\vbox{\baselineskip0pt \lineskip-.5pt
x  \ialign{$\mxth#1\hfil##\hfil$\crcr#2\crcr\sim\crcr}}}

\renewcommand{\a}{\alpha}
\renewcommand{\b}{\beta}

\renewcommand{\d}{\delta}
\newcommand{\pa}{\partial}
\newcommand{\g}{\gamma}
\newcommand{\G}{\Gamma}
\newcommand{\e}{\epsilon}
\newcommand{\z}{\zeta}
\renewcommand{\l}{\lambda}
\renewcommand{\L}{\Lambda}
\newcommand{\m}{\mu}
\newcommand{\n}{\nu}

\newcommand{\s}{\sigma}

\newcommand{\Y}{\Upsilon}
\renewcommand{\o}{\omega}

\newcommand{\Ka}{{K\"ahler}}
\renewcommand{\O}{{\Omega}}

\def\be{\begin{equation}}
\def\ee{\end{equation}}
\def\bea{\begin{eqnarray}}
\def\eea{\end{eqnarray}}

\newcommand{\ft}[2]{{\textstyle\frac{#1}{#2}}}

\newcommand{\eqn}[1]{(\ref{#1})}

\begin{document}
\begin{titlepage}
\begin{center}
\hfill ITP-UU-01/02  \\
\hfill SPIN-01/02  \\
\hfill YITP-01/04  \\
\hfill {\tt hep-th/0101161}\\
\vskip 20mm

{\Large {\bf Hypermultiplets, Hyperk\"ahler Cones }}\\[2mm]
{\Large {\bf and Quaternion-\Ka\ Geometry }}

\vskip 10mm

{\bf Bernard de Wit$^{a}$, Martin Ro\v{c}ek$^b$  and
Stefan Vandoren$^b$ }

\vskip 4mm

$^a${\em Institute for Theoretical Physics} and {\em Spinoza
Institute}\\
{\em Utrecht University, Utrecht, The Netherlands}\\
{\tt  B.deWit@phys.uu.nl} \\[2mm]
$^b${\em C.N. Yang Institute for Theoretical Physics}\\
{\em SUNY, Stony Brook, NY 11794-3840, USA}\\
{\tt rocek@insti.physics.sunysb.edu}\\
{\tt vandoren@insti.physics.sunysb.edu}

\vskip 6mm

\end{center}

\vskip .2in

\begin{center} {\bf ABSTRACT } \end{center}
\begin{quotation}\noindent
We study hyperk\"ahler cones and their corresponding quaternion-\Ka\ spaces.
We present a classification of $4(n-1)$-dimensional quaternion-\Ka\ spaces
with $n$ abelian quaternionic isometries, based on dualizing superconformal
tensor multiplets. These manifolds characterize the geometry of the
hypermultiplet sector of classical and perturbative moduli spaces of 
type-II strings compactified on a Calabi-Yau manifold. As an example of 
our construction, we study the universal hypermultiplet in detail, and 
give three inequivalent tensor multiplet descriptions. We also comment on the
construction of quaternion-\Ka\ manifolds that may describe instanton
corrections to the moduli space.

\end{quotation}

\vfill
\flushleft{\today}

\end{titlepage}

\eject

\section{Introduction}
Hyperk\"ahler and quaternion-K\"ahler manifolds, whose real dimensions are
multiples of four, appear in various contexts in field and string theory.  By
definition, hyperk\"ahler spaces of dimension $4n$ have a holonomy group
contained in Sp($n$); they are Ricci flat. Examples of hyperk\"ahler spaces
are the moduli spaces of magnetic monopoles (such as the Taub-Nut and
Atiyah-Hitchin manifolds \cite{AH}), or the moduli spaces of Yang-Mills
instantons in flat space, as described by the ADHM construction \cite{ADHM}.
Other examples of hyperk\"ahler spaces are four-dimensional gravitational
instantons (such as the Eguchi-Hanson metric) and K3 surfaces. Furthermore,
in rigidly  supersymmetric sigma models with 8 supercharges the scalar
fields are known to parametrize a hyperk\"ahler target space\footnote{To be
precise, this is true only for spacetime (or worldsheet, as opposed to
target-space) dimension greater than two; in two dimensions, target-space
torsion can modify the geometry.} \cite{AG-Fr,DDKV}. In four spacetime
dimensions such a sigma model has $N=2$ supersymmetry and is based on $n$
hypermultiplets, each consisting of four real scalars and two Majorana
spinor fields. In what follows, we work in four spacetime dimensions, though
many of our conclusions concern the geometry of the target-space, and are
independent of the spacetime dimension.

When the supersymmetry is realized locally, the hypermultiplets couple
to $N=2$ supergravity and the target space becomes quaternion-\Ka\
\cite{BagWit}. Quaternion-\Ka\  spaces of dimension $4n$ have a holonomy
group contained in ${\rm Sp}(n)\cdot {\rm Sp}(1)$, with nontrivial Sp(1)
holonomy; they are Einstein spaces. The simplest (compact) four-dimensional
quaternion-\Ka\ spaces are the sphere $S^4$ and  the complex projective
space $CP^2= {\rm SU}(3)/{\rm U}(2)$. Quaternion-K\"ahler spaces appear as
(part of) the moduli space  of Calabi-Yau manifolds. Therefore they appear as
hypermultiplet target spaces in  the low-energy effective action for
type-II superstrings compactified on a Calabi-Yau manifold. Classically
these moduli spaces are known \cite{cfg}, but little is understood about
the perturbative \cite{quat-quantum} and non-perturbative \cite{nonpert}
corrections to them.  One reason is that our knowledge of quaternion-\Ka\
geometry is limited, and no convenient formulation is known that allows one
to address these questions effectively.

{From} the $N=2$ superconformal multiplet calculus \cite{DVV} it is clear
that there exists a relation between quaternion-\Ka\ manifolds and certain
hyperk\"ahler spaces. These are {\it hyperk\"ahler
cones}\footnote{In \cite{DWKV} these spaces were called
{\it special} hyperk\"ahler, but to avoid confusion with
hyperk\"ahler manifolds related by the {\bf c}-map to {\it special} \Ka\
geometries, we have changed our nomenclature and call them hyperk\"ahler
cones (HKC). We thank A. Van Proeyen for his encouragement on this
issue.} and correspond to field theories that are invariant under rigid $N=2$
superconformal symmetry
\cite{dubna,DWKV}. Such a cone (relevant, {\it e.g.}, for the moduli space
of  Yang-Mills instantons) has a homothetic Killing vector and three complex
structures which rotate isometrically under the group Sp(1)
\cite{Swann,Galicki}. It is a cone over a $(4n-3)$-dimensional 3-Sasakian
manifold, which in turn is an Sp(1) fibration of a $(4n-4)$-dimensional
quaternion-\Ka\ space. The quaternion-\Ka\ manifold is the $N=2$ {\em
superconformal quotient} of the hyperk\"ahler manifold. There is a
one-to-one relation between quaternion-\Ka\ spaces and hyperk\"ahler cones;
it has been extended to the case with torsion in \cite{hop}.

In this paper we give a detailed description of the $N=2$ superconformal
quotient and study other aspects of hyperk\"ahler cones, such as their
isometry structure and their dual description in terms of rigidly
superconformally invariant actions of tensor multiplets. The superconformal
quotient can be performed in two steps: First, one descends from the
hyperk\"ahler cone to the {\em twistor space} \cite{Salamon,Swann}, which
is a \Ka\ quotient of the hyperk\"ahler cone and is an $S^2$ fibration of
the underlying quaternion-\Ka\ space. The twistor space plays an
intermediate role in the explicit construction of the quaternion-\Ka\
space. Then one projects from the twistor space down to the
quaternion-\Ka\ space, which can be done by imposing a gauge condition.
Triholomorphic isometries of a hyperk\"ahler cone lead to
quaternionic isometries (isometries that rotate the quaternionic
structure) on the corresponding quaternion-\Ka\ space.

Any $N=2$ superconformal theory of $n$ tensor multiplets has a dual
description in terms of $n$ hypermultiplets whose target space is a
hyperk\"ahler cone with $n$ abelian triholomorphic isometries. For such a
space there is a systematic representation in terms of a homogeneous
function of the tensor multiplet scalars and an auxiliary complex variable
$\xi$ integrated along a closed loop in the complex $\xi$-plane
\cite{KLR,Hitchin}. For the hyperk\"ahler cones, this function is
remarkably simple (as compared to the functions that appear in the
Lagrangian of the tensor multiplets), and therefore provides
an effective way of studying these spaces. Performing their $N=2$
superconformal quotient yields quaternion-\Ka\ spaces of real dimension
$4n-4$ with $n$ abelian quaternionic isometries. Precisely these manifolds
occur in the low-energy effective actions of type-II superstrings on
Calabi-Yau manifolds, because the perturbative corrections to the
hypermultiplet moduli spaces respect certain Peccei-Quinn isometries and
hence fall into this class. As all quaternion-\Ka\ spaces with
$n$ abelian quaternionic isometries can be constructed in terms of tensor
multiplets, one obtains a general classification of these spaces in terms of
the homogeneous functions mentioned above. However, the homogeneous
functions fall into equivalence classes, so that different functions can
lead to the same hyperk\"ahler cone or corresponding quaternion-\Ka\ space.
This aspect is studied at the end of this paper.

The paper is organized as follows. In section~2 we review
aspects of hyperk\"ahler cones and quaternion-K\"ahler geometry.
This section summarizes some of the results of \cite{DWKV} and
emphasizes the relevance of hyperk\"ahler quotients and superconformal
quotients.
In section~3 we construct the twistor space, which
is an Einstein-\Ka\ manifold of dimension $4n-2$ and plays an
intermediate role in the description of the superconformal
quotient. We also discuss the isometries of the twistor space
that descend from triholomorphic isometries of the corresponding
hyperk\"ahler cone.
In section~4 we derive the $(4n-4)$-dimensional quaternion-\Ka\
geometry that corresponds to the twistor space and
give explicit formulae for the quaternion-\Ka\ metric and quaternionic
structure in terms of twistor space quantities. We also exhibit how
hyperk\"ahler cone and twistor space isometries descend to the
quaternion-K\"ahler manifold. In section~5 we construct general Lagrangians
for $n$ $N=2$ tensor supermultiplets that are rigidly superconformally
invariant. These actions are encoded in homogeneous functions, which, as
mentioned above, have a contour integral representation. The tensor fields
can be dualized so that one obtains a field theory of $n$ hypermultiplets
whose target space is a hyperk\"ahler cone with $n$ abelian triholomorphic
isometries. In section~6 we study the corresponding quaternion-\Ka\ spaces
with $n$ abelian quaternionic isometries and explain their classification.
In section~7 we discuss the geometry of unitary Wolf spaces, and
specifically the universal hypermultiplet from the various points of view
developed in this paper. We describe them as coset spaces, as
combined hyperk\"ahler and superconformal quotients, and in terms of
tensor multiplets. We note the appearence of inequivalent tensor
multiplet descriptions. In section~8 we discuss our results from the point
of view of applications and mention open problems and future perspectives.

We have added two appendices. In appendix~A we discuss and derive the
restrictions on functions that encode the superconformally invariant
tensor multiplet Lagrangians. In appendix~B we present a self-contained
description of projective superspace (from which the contour integral
representation arises naturally), discuss gauging triholomorphic isometries,
and give applications to hyperk\"ahler quotients and tensor multiplet
dualities.

A brief summary of our main results will appear in \cite{dwrv}.

\section{Preliminaries}
\setcounter{equation}{0}
In this section, we briefly review properties of  hyperk\"ahler cones
and discuss \Ka\ and superconformal quaternion-\Ka\ quotients.

\subsection{Hyperk\"ahler cones}
Hyperk\"ahler cones \cite{dubna,DWKV,Swann,Galicki} have a
homothetic conformal Killing vector $\chi^A$:
\be
D_A\chi^B = \d_A{}^B\,, \qquad (A,B = 1,\ldots, 4n)\ . \label{homothety}
\ee
Hence the hyperk\"ahler cone can be characterized by
a hyperk\"ahler potential $\chi$, which serves as a \Ka\ potential for
each of the three complex structures. This potential can be expressed
in terms of the HKC  metric and $\chi^A$ as
\be
\chi = \ft12 \chi^A g_{AB} \chi^B\ . \label{hyperK-potential}
\ee
The derivative of the hyperk\"ahler potential is (locally) equal to the
homothetic one-form,
\be
\chi_A = \pa_A\chi\ .
\label{homothetic}
\ee

The three covariantly constant complex structures of the hyperk\"ahler cone
are denoted by $\vec J{}^{A}{}_{\!B}$. They are hermitean, {\it i.e.}, $
{\vec \O}_{AC}\equiv  g_{AB}\vec J{}^B{}_{\!C}$ is antisymmetric, and they
obey the algebra of the quaternions:
\be
J^{\Pi}J^\Sigma=-g^{\Pi\Sigma}+\varepsilon^{\Pi\Sigma}
{}_{\Lambda}\,J^\Lambda\ ,
\ee
which in a complex basis with components $J^3$ and $J^\pm =
\ft12(J^1\mp i J^2)$, becomes
\bea
J^\pm \,J^3 = \pm i\,J^\pm \ ,\quad (J^3)^2 = -{\bf 1}\ ,
\quad
(J^\pm)^2 = 0\ ,\quad J^+\,J^- = -\ft12 ({\bf1} +i J^3)\ .\label{JJ}
\eea

Hyperk\"ahler cones have an Sp(1) isometry whose Killing
vectors are
\be
\vec k{}^A = \vec J{}^{A}{}_{\!B}\, \chi^{B}\ .
\label{su(2)-vector}
\ee
To show that these are indeed Killing vectors, we note that
\be
D_A \vec k{}_B = - \vec \Omega_{AB}\ ,
\label{D-su(2)}
\ee
by virtue of \eqn{homothety}.
The Sp(1) isometries are not triholomorphic,
{\it i.e.}, they do not leave the complex structures
invariant. Instead the complex structures
rotate under Sp(1) as
\be
{\cal L}_{\vec \epsilon\cdot \vec k} J^{\Lambda\,A}{}_{\! B}\equiv
\vec \epsilon \cdot \Big
( \vec k^C\,\pa_C J^{\L\,A}{}_{\! B} - \pa_C\vec k^A\, J^{\L\,C}{}_{\!
B}  + \pa_B\vec k^C \, J^{\L\,A}{}_{\! C} \Big)
    = 2\varepsilon^{\L\Pi}{}_\Sigma\,\e_\Pi\,J^{\Sigma\, A}{}_B\ ,
\ee
which becomes
\be
{\cal L}_{\vec \epsilon\cdot \vec k} J^{\pm} =
\pm 2i (\e^3 \,J^\pm - \e^\pm \,J^3)\ ,\qquad
{\cal L}_{\vec \epsilon\cdot \vec k} J^3  =
4i (\e^+\, J^- - \e^-\,J^+)   \ ,\label{rot-J}
\ee
in the complex basis.
Here $\vec \e\cdot \vec k = \e^3k^3 + 2 (\e^+k^-+ \e^-k^+)$.

The hyperk\"ahler potential $\chi$ is Sp(1) invariant.
The four vectors associated with the homothetic conformal Killing
vector, $\chi^A$, and the three Sp(1) Killing vectors,  $k^{3A}$ and
$k^{\pm A}$, define a subspace that is locally flat, {\it i.e.}, the Riemann
tensor vanishes when contracted with any of these four vectors. We recall
from \cite{DWKV} that these four vectors are orthogonal
(cf. \eqn{su(2)-vector}) and normalized according to
\be
\label{norms}
\chi^{A} \, \chi{}_A = k^{3A}k^3_A = 2\, k^{+A}k_A^- = 2\,\chi\ ,
\ee
with all other inner products vanishing.

Spaces with a homothety can always be described as a
cone. This becomes manifest when decomposing the coordinates $\phi^A$
into coordinates tangential and orthogonal to the $(4n-1)$-dimensional
hypersurface defined by setting $\chi$ to a constant. The line element
can then be written in the form \cite{GibbonsRych},
\be
{\rm d}s^2 = {{\rm d}\chi^2\over 2 \chi} + \chi\,
h_{\hat A\hat B}(x) \,{\rm d}x^{\hat A}\,{\rm d}x^{\hat B} \ ,
\label{conemetr}
\ee
where the $x^{\hat A}$ are the coordinates associated with the
hypersurface. In the present case this hypersurface is a
3-Sasakian space\footnote{For a review on 3-Sasakian manifolds
we refer the reader to \cite{sasaki}; note that ${\bf S}^{4n-1}$ is in
general {\em not} the sphere $S^{4n-1}$.}
${\bf S}^{4n-1}$, and the hyperk\"ahler  space is therefore a cone
over ${\bf S}^{4n-1}$. As is well known from the mathematics
literature \cite{Swann}, the 3-Sasakian space is an Sp(1) fibration of
a $(4n-4)$-dimensional quaternion-\Ka\ manifold ${\bf
Q}^{4n-4}$. Hence the manifold can be written as
$R^+ \times[{\rm Sp(1)} \rightarrow {\bf S}^{4n-1}\rightarrow
{\bf Q}^{4n-4}]$. Another relevant fibration of the quaternion-\Ka\
manifold is the twistor space $\cal Z$, which is a $(4n-2)$-dimensional
Einstein-\Ka\ manifold \cite{Salamon,Swann}. In the next subsection and in
section 3, we explicitly construct this twistor space from the HKC
geometry.

Some of our results of sections 3 and 4 are illustrated in a few
examples, based on a $(4n)$-dimensional flat
space, which is obviously a hyperk\"ahler cone. In view of supergravity
applications we allow for pseudo-Riemannian metrics. We use
complex coordinates\footnote{We write capital
letters $A,B, \ldots=1,2,\ldots,4n$ for real
coordinates and small letters $a,b,\ldots= 1,\ldots,2n$ for
holomorphic coordinates.} $z^a$, with $a=1,\ldots,2n$, and corresponding
anti-holomorphic ones $\bar z^{\bar a}$, with a metric $\eta_{a\bar b}$
that can be chosen diagonal with even numbers of positive and negative
eigenvalues. The coordinate basis is chosen such
that $J^{3\,a}{}_{\!b}= i \d^a{}_{\!b}$. The two other complex structures,
$J^+$ and $J^-$, are associated with a holomorphic and an
anti-holomorphic two-form, denoted by $\Omega$ and $\bar\Omega$,
respectively. The various quantities of interest for flat
space can then be defined as
\bea
\chi(z,\bar z)&=& \eta_{a\bar b} \,z^a\bar z^{\bar b} \ ,\nonumber\\
\Omega^3&=& -i\,\eta_{a\bar b}\, {\rm d} z^a \wedge {\rm d} \bar z^{\bar b}
\ ,\nonumber\\
\Omega &=& \ft12 \Omega_{ab} \;{\rm d} z^a \wedge {\rm d} z^{b} \ ,
\nonumber \\
\bar \Omega &=& \ft12  \bar \Omega_{\bar a\bar b} \;{\rm d} \bar
z^{\bar a} \wedge {\rm d} \bar z^{\bar b} \ .
\label{pseudoRiemannian}
\eea
The tensors $\Omega_{ab}$ are constant skew-symmetric  and satisfy
$\Omega_{ab}\,\bar\Omega_{\bar a\bar b} \,\eta^{\bar a b} =
-\eta_{a\bar b}$, where $\eta^{\bar a b}$ is the inverse metric.

\subsection{K\"ahler and $N=2$ superconformal quotients}

The metrics of the twistor space and the quaternion-\Ka\
space can be expressed directly in terms of the HKC
metric by performing appropriate quotients. The resulting metric
is horizontal to a certain subspace but does not come equipped with
unique canonical coordinates. A choice of coordinates can be found by
imposing gauge conditions associated with the isometries upon which the
quotient is based. These quotients are at the heart of the superconformal
multiplet calculus of supergravity \cite{DVV}.

In general, when the hyperk\"ahler cone has an isometry with a
Killing vector $k^A$ that commutes with the dilatations, $k^A\chi_A=0$;
this implies that $\chi$ and $\chi_A$ are invariant. Hence
\be
\label{3sas-metr}
\frac{1}{\chi}\Big( g_{AB}-\frac{1}{2\chi}\chi_A \chi_B \Big)
\ee
is preserved by $k^A$. Imposing the constraint $\chi = \rm constant$,
it follows from \eqn{conemetr} that this is the 3-Sasakian metric.

The quotient metric is well known; physicists find it by constructing a
$\sigma$-model with \eqn{3sas-metr} as the metric and gauging the $k^A$
isometry by covariantizing spacetime derivatives: $\pa_\mu\phi^A \to D_\mu
\phi^A=\pa_\mu\phi^A -A_\mu k^A$. The gauge field $A_\mu$ can then be
eliminated by its field equation:
\be
A_\mu=\frac{1}{k^Bk_B}\,k_A \pa_\mu \phi^A\ .
\ee
Substituting this result into the $\sigma$-model action leaves the gauge
invariance unaffected and has the effect of changing the metric
\eqn{3sas-metr} into
\be
G_{AB}=\frac{1}{\chi}\left(g_{AB}-\frac{1}{2\chi}\, \chi_A\chi_B -
\frac{1}{k^Ck_C}\,k_A k_B\right)\ .
\label{horizontal}
\ee
Observe that this metric is horizontal in the sense that its
contraction with $\chi^A$ and $k^A$ vanishes. In the horizontal
subspace, it is nondegenerate and precisely the quotient metric.

The twistor space is the quotient with respect to the $k^3$
isometry, and has the quotient metric:
\be
G_{AB}=\frac{1}{\chi}\left(g_{AB}-\frac{1}{2\chi}\Big[ \chi_A\chi_B +
k^3_A k^3_B\Big]\right)\ ,
\label{twistorhorizontal}
\ee
where we have used \eqn{norms}. Because of \eqn{su(2)-vector}, $k^3$
is holomorphic with respect to $J^3$, and this is a standard \Ka\
quotient \cite{KQ}. The moment map of the holomorphic $k^3$ isometry is the
hyperk\"ahler potential $\chi$.

To obtain the quaternion-\Ka\  space, the quotient is taken with respect
to the full Sp(1) isometry group. Hence one introduces gauge fields
${\vec A}_\mu$ and covariantizes the derivatives, $D_\mu
\phi^A=\pa_\mu \phi^A -{\vec A}_\mu\cdot {\vec k}^A$. The field
equations now yield
\be
{\vec A}_\mu=\ft12 \chi^{-1}{\vec k}_A\,  \pa_\mu \phi^A\ .
\label{sp1-gauge}
\ee
Substituting this result back into the action (which leaves the Sp(1)
gauge invariance unaffected) leads to a new metric orthogonal to all
four vectors $\chi^A$ and $\vec k^A$. This is the horizontal metric
of \cite{DWKV}:
\be
G_{AB}=\frac{1}{\chi}\left(g_{AB}-\frac{1}{2\chi}\Big[ \chi_A\chi_B +
{\vec k}_A\cdot{\vec k}_B\Big] \right)\ .
\label{quaternionhorizontal}
\ee
However, as the Sp(1) isometries are not triholomorphic in the
hyperk\"ahler cone, the above quotient not a standard hyperk\"ahler
quotient \cite{Hitchin}; such quotients we call {\it $N=2$
superconformal quotients}. As a result, the metric
\eqn{quaternionhorizontal} is no longer hyperk\"ahler but
rather quaternion-K\"ahler.

It is always possible to choose a coordinate along the $k^3$ Killing vector;
the metric of the twistor space ${\cal Z}$ is evidently independent of this
coordinate. Consequently, the HKC metric naturally projects to the
twistor space metric without the need for imposing gauge conditions. The
situation regarding the quaternion-\Ka\ metric is different in this respect.
Here we project out an $S^2\cong {\rm Sp}(1)/{\rm U}(1)$ from the
twistor space. Because there are no corresponding Killing vectors, one
has to impose appropriate gauge conditions. This is discussed in
section 4.

Similarly, the three quaternionic two-forms can be constructed
by projecting the HKC complex structures onto the horizontal space,
\be
\label{QK-cc}
{\vec {\cal Q}}_{AB}=G_{AC}{\vec J}^C{}_B\ .
\ee
These tensors satisfy the quaternionic algebra relations \cite{DWKV}
\bea
{\cal Q}^{\pm}_{AC}\,\chi\,g^{CD}{\cal Q}^3_{DB}=\pm i {\cal Q}^{\pm}_{AB}\ ,
&\quad&
{\cal Q}^+_{AC}\,\chi\,g^{CD}{\cal Q}^-_{DB}=-\ft12(G_{AB}+i{\cal Q}^3_{AB})
\ ,\nonumber\\[2mm]
{\cal Q}^3_{AC}\,\chi\,g^{CD}{\cal Q}^3_{DB}=-G_{AB} \ ,&\quad&
{\cal Q}^{\pm}_{AC}\,\chi\,g^{CD}{\cal Q}^{\pm}_{DB}=0\ .
\eea
Even though $\chi\,g^{AB}$ is not horizontal, it acts as an
inverse metric on the horizontal
subspace because it satisfies $ G_{AC}\,\chi\,g^{CD}G_{DB}=G_{AB}$.

Quaternion-\Ka\ manifolds have non-trivial Sp(1) holonomy.
In \cite{DWKV} the Sp(1) connection was given in terms of
the Sp(1) Killing vectors of the hyperk\"ahler cone,
\be
\label{Sp1-conn}
{\vec{\cal V}}_{A}=\chi^{-1}\,\vec{k}_A\ .
\ee
This vector is invariant under the homothety and rotates under the
Sp(1) isometries as a vector. Up to normalization, its pull-back is
the gauge field \eqn{sp1-gauge}.
The curvature associated with this connection is
proportional to the two-forms \eqn{QK-cc}, as is required for a
quaternion-\Ka\ geometry. We return to this and related points in
section~4.

\section{Reduction to the twistor space ${\cal Z}$}
\setcounter{equation}{0}
Consider a $4n$-dimensional hyperk\"ahler cone with hyperk\"ahler potential
$\chi$ parametrized by $2n$ holomorphic coordinates $z^a$.
Note that \eqn{homothety} implies that the homothetic conformal Killing
vector has holomorphic and anti-holomorphic components
$\chi^a(z)$ and $\chi^{\bar a}(\bar z)$, respectively:
\be
\chi(z,\bar z)=\chi^a(z)\,\chi_a(z,\bar z)=\chi^{\bar a}(\bar
z)\,\chi_{\bar a}(z,\bar z)\ ,\label{chi}
\ee
where, {\it e.g.},  $\chi_a(z,\bar z)=g_{a{\bar b}}(z,\bar z)\,\chi^{\bar b}
(\bar z)$, and the metric is
\be
g_{a\bar b} (z,\bar z) = \pa_a\pa_{\bar b} \chi(z,\bar z)\ .
\ee

The holomorphic vector field $\chi^a$ and its complex
conjugate can be used to define new complex coordinates.
One special coordinate is denoted by $z$ (not to be confused
with the $z^a$) and the $2n-1$ remaining coordinates by $u^i$; the
precise definition of the $u^i$ is of no concern. The $u^i$ turn out
to parametrize an Einstein-\Ka\ manifold, the twistor space
$\cal Z$ \cite{Salamon}. This space is an $S^2$ fibration of an
underlying quaternion-\Ka\ manifold that we discuss in section 4. The
coordinate $z$ is defined (up to a special class of holomorphic
diffeomorphisms, see below) by
\be
\chi^a(z,u)\,{\partial\over\pa z^a} \equiv {\partial\over\partial z} \ ,
\label{def-z}
\ee
so that, upon using \eqn{homothetic}, \eqn{chi} may be regarded as a
first-order differential equation for $\chi$
which  determines its dependence on the new coordinates $z$ and $\bar
z$. The result is that $\chi(z,\bar z,u,\bar u)$ can be written as
\be
\label{chi-K}
\chi(z,\bar z; u,\bar u)={\rm e}^{z+{\bar z}+ K(u,{\bar u})}\ .
\ee
The function $K(u,{\bar u})$ is the \Ka\ potential of the \Ka\ quotient
of the hyperk\"ahler cone with respect to the U(1) isometry generated by $k^3$
and the compatible \Ka\ structure $\O^3$ \cite{KQ,Hitchin}.
This quotient is the twistor space ${\cal Z}$; it is Einstein-\Ka\
with metric $K_{i\bar \jmath}(u,\bar u)$, and has complex dimension $2n-1$.

Observe that \Ka\ transformations for this twistor space,
$K(u,\bar u) \to K(u,\bar u) + f(u) + \bar f(\bar u)$, can be compensated
by corresponding coordinate changes $z\to z - f(u)$, and hence the coordinate
$z$ is defined only modulo this ambiguity. In contrast,
the hyperk\"ahler potential $\chi$ in general cannot be redefined by means
of a \Ka\ transformation because \eqn{homothety} fixes this freedom.

The HKC metric in the new coordinates $(u^i,z)$ is
\be
g_{a{\bar b}}=\partial_a\partial_{\bar b}\chi=\chi
\pmatrix{ K_{i{\bar \jmath}}+K_iK_{\bar \jmath}
 & K_i \cr \noalign{\vskip 3mm} K_{\bar \jmath}  & 1}\ ,
\ee
where we use the notation $K_i=\partial_i K(u,\bar u)$,  etc..
The HKC  line element takes the form
\be
{\rm d}s^2=\chi\Big[K_{i{\bar \jmath}}\,{\rm d}u^i
{\rm d}{\bar u}^{\bar \jmath}+
({\rm d}z+K_i\,{\rm d}u^i)
({\rm d}{\bar z}+K_{\bar \jmath}\,{\rm d}{\bar
u}^{\bar \jmath}) \Big]\ .
\ee
The inverse metric can be computed and equals
\be
g^{{\bar a}b}=\chi^{-1}\pmatrix{ K^{{\bar \imath}j} & -K^{\bar \imath}
\cr \noalign{\vskip 3mm}  -K^j &
1+K^lK_l}\ ,
\ee
where $K^{{\bar \imath}j}(u,\bar u)$ denotes the inverse of
$K_{i{\bar \jmath }}$. In the following we use this metric to raise
and lower indices as in $K^{\bar \imath}=K^{{\bar \imath}j}K_j$.

The HKC Christoffel symbols $\G_{ab}{}^c=(\partial_a g_{b{\bar d
}})\,g^{{\bar d}c}$ are
\bea
\G_{za}{}^b&=&\delta_a{}^b \ ,\nonumber\\
\G_{ij}{}^z&=&K_{ij}-K_iK_j-\gamma_{ij}{}^k\,K_k \ ,\nonumber \\
\G_{ij}{}^k&=&\gamma_{ij}{}^k+2\,K_{(i}\delta_{j)}{}^k\ ,
\label{C-connections}
\eea
where $\gamma_{ij}{}^k$ is the Christoffel connection for the
twistor space ${\cal Z}$ and (anti) symmetrization is always done
with weight one, e.g. $(ij)=\ft12(ij+ji)$.

Similarly, we compute the HKC  Riemann tensor
$R_{\bar a bc}{}^d= \partial_{\bar a}\G_{bc}{}^d$;
as the connection is independent of $z$ and ${\bar z}$, $R_{\bar a z
b}{}^c=R_{{\bar z} ab}{}^c=R_{{\bar a}b z}{}^c=0$, {\it i.e.}, the
curvature vanishes when contracted with the homothetic Killing vector, as
claimed in the previous section. The remaining components are
\begin{equation}
R_{{\bar \imath}{ j}k}{}^z=-{\cal R}_{{\bar \imath}{j}k}{}^l\,K_l-
2\,K_{(j}K_{k) {\bar \imath}}\ ,
\qquad
R_{{\bar \imath}{ j}k}{}^l={\cal R}_{{\bar \imath}{ j}k}{}^l+
2\delta_{(j}{}^lK_{k){\bar \imath}}\ ,
\end{equation}
where ${\cal R}_{{\bar \imath} j k}{}^l$ is the Riemann tensor
of the twistor space $\cal Z$.

Being hyperk\"ahler, the HKC is Ricci-flat, and hence the twistor space
${\cal Z}$ is Einstein with positive cosmological constant $2n$:
\be
{\cal R}_{i{\bar \jmath}}=-2n\,K_{i{\bar \jmath}}\ .\label{EK}
\ee
For any \Ka\ manifold $R_{a\bar b}=\pa_a\pa_{\bar b}\ln\det (g_{c{\bar
d}})$; for the HKC metric we have explicitly
\be
\det (g_{a{\bar b}})=\chi^{2n}\,\det (K_{i{\bar \jmath}})\ ,
\ee
which must therefore be a product of a holomorphic and an anti-holomorphic
function. Hence
\be
\det (g_{a{\bar b}})= \vert {\rm e}^{2 z + f(u)}\vert^{2n}
\,, \qquad \det(K_{i{\bar \jmath}})=
\vert {\rm e}^{f(u)}\vert^{2n}
\,{\rm e}^{-2nK(u,{\bar u})}\ ,
\label{K-equation}
\ee
where $f(u)$ is the arbitrary holomorphic function that can be
absorbed into the \Ka\ potential by performing a \Ka\
transformation $K(u,\bar u) \to K(u,\bar u) + f(u) + \bar f(\bar
u)$ on ${\cal Z}$. This is of course consistent with \eqn{EK}:
\be
{\cal R}_{i{\bar \jmath}}
=\partial_i\partial_{\bar \jmath}\,\ln[\det K_{k{\bar l }}]  =
-2n\,K_{i{\bar \jmath}}\ .
\ee

We now examine how the three \Ka\ forms ${\vec \O}$
of the hyperk\"ahler cone descend to ${\cal Z}$.
Because $\O^+$ is covariantly constant, it depends only on holomorphic
coordinates, and we denote it by $\Omega_{ab}(u,z)$; similarly,
$\O^{ab}\equiv {\bar\O}_{\bar c \bar d} g^{\bar c a}g^{\bar d b}$, which
obeys $\O_{ac}\O^{cb}=J^{+\,c}{}_{\!a}J^{-\,b}{}_{\!c}=-\delta_a^{\,b}$, is
also holomorphic. Computing the covariant derivative with respect to $z$,
we find, in the coordinates $(u^i,z)$,
\bea
\Omega_{ab}(u,z)&=& {\rm e}^{2z} \pmatrix{\omega_{ij}(u) & X_i(u)
\cr \noalign{\vskip3mm} -X_j(u)&0}\ , \nonumber\\[3mm]
\Omega^{ab}(u,z)&=& {\rm e}^{-2z}\pmatrix{\hat\omega^{ij}(u) &
Y^i(u)\cr \noalign{\vskip3mm}-Y^j(u)&0}\, .
\eea
We can express $\hat \omega^{ij}$ and $Y^i$ in terms of $X_i$,
$\omega_{ij}$ and the \Ka\ potential and its derivatives,
\bea
\hat\omega^{ij}(u) &=& \Big[ \omega^{ij} +2\, K^{[i}X^{j]} \Big]
\,{\rm e}^{-2K}\ ,\nonumber \\
Y^i(u) &=& \Big[(1+K_jK^j) X^i -K^i\,X^j K_j -
\omega^{ij}K_j  \Big] \,{\rm e}^{-2K}\nonumber \\
&=&X^i\,{\rm e}^{-2K} -{\hat \o}^{ij}K_j\ ,
          \label{Y-hat-omega}
\eea
where we raise and lower indices with $K^{\bar \imath j}$ and $K_{i\bar
\jmath}$, $X^i= K^{i\bar\jmath}
\,X_{\bar \jmath}$, $X_{\bar\imath} = (X_i)^\ast$, and similarly
for $\omega^{ij}$. Though it is not manifest, the right-hand sides of
\eqn{Y-hat-omega} are nonetheless holomorphic.

The relation $\Omega_{ac}(u,z)\,\Omega^{cb}(u,z)= -
\d_a^{\,b}$ implies the following identities,
\bea
X_i\,Y^i &=& 1\ ,\nonumber\\
\omega_{ij}\,Y^j &=&0\ , \nonumber \\
\hat \omega^{ij}\,X_j &=& 0 \ ,\nonumber\\
\hat \omega^{ik} \omega_{kj}  &=& -\d_j^i + Y^i\,X_j\ .
\label{first-id}
\eea
The first two equations imply that ${\cal L}_Y X_i =
0$, whereas the second and third display the null vectors of the
odd-dimensional antisymmetric tensors $\omega_{ij}$ and $\hat\omega^{ij}$.
Combining the above results with \eqn{Y-hat-omega} leads to additional
identities, such as
\bea
X_iX^i &=&{\rm e}^{2K} \ ,\nonumber\\
\omega_{ij} X^j &=& X_i \, K_jX^j  - K_i\,{\rm e}^{2K}  \ ,\nonumber\\
Y^iY_i &=& (1+ K_iK^i) \,{\rm e}^{-2K} - \vert Y^iK_i\vert^2
\ ,\nonumber\\
X^iK_i&=&Y^iK_i \,{\rm e}^{2K}\ .
\label{add-id}
\eea
The structure of the HKC does not imply any constraints that do not follow
from those found above.

In these coordinates, the homothetic and Sp(1) Killing vectors are
\bea
&&\chi^a = -i k^{3a} = (0,\ldots,0,1) \ , \qquad \chi_a = i k^3_a= \pa_a
\chi =\chi (K_i,1) \ , \nonumber\\
&&k^+_a = \Omega_{az} = {\rm e}^{2z} (X_i,0)\ ,\qquad\qquad k^-_{\bar a} =
\Omega_{\bar a\bar z}\ ,
\eea
and we may explicitly verify \eqn{twistorhorizontal}:
\be
K_{i\bar\jmath} = {1\over \chi} \left( g_{i\bar\jmath} - {1\over 2\chi}
\Big[ \chi_i\,\chi_{\bar\jmath} + k^3_i\,k^3_{\bar\jmath}\Big] \right)\ .
\ee

Observe that $k^+_a$ is holomorphic and $k^-_{\bar a}$ is
anti-holomorphic; raising the index of the latter one finds
\be
\label{k-upper}
k^{-a} = \chi^{-1}  {\rm e}^{2 \bar z} \,(X^i,- X^jK_j)\ ,
\ee
where we made use of the previous identities. One can verify that
the orthogonality conditions of these four vectors are indeed
satisfied, {\it e.g.},
$k^{-a}k^+_a=\chi$.
Furthermore one can verify that $D_ak^+_b = - \Omega_{ab}$ and
$D_{\bar a} k^-_{\bar b} =- {\bar \Omega}_{\bar a\bar b}$ as specified by
\eqn{D-su(2)}. This leads directly to
\be
D_iX_j =- \o_{ij}+ 2\,K_{(i}X_{j)}\ ,\label{DX}
\ee
where $D_i$ contains only the Christoffel connection $\gamma_{ij}{}^k$
of ${\cal Z}$; the second term is due to the extra term in the
hyperk\"ahler connection $\G_{ij}{}^k$. Hence it follows that
$\omega_{ij}$ is (locally) exact. The pair $X,\o$ is a contact structure.
The result \eqn{DX} is also required by covariant constancy of $\Omega_{ab}$.
However, though \eqn{DX} implies $D_{(i} ({\rm e}^{-2K}X_{j)})= 0$, ${\rm
e}^{-2K}X_i$ is not a Killing vector of the twistor space, as
$\pa_{(\bar\imath} ({\rm e}^{-2K}X_{j)}) \ne 0$. We return to this point in
the next section where we discuss the role played by $X^i$ for the
quaternion-\Ka\ space.

All hyperk\"ahler cones have a homothety and Sp(1) isometries; in some
cases, they may have additional isometries. A triholomorphic isometry leaves
the complex structures invariant. Not
all HKC isometries descend to isometries of the twistor
space; for example, the $k^{\pm}$ isometries are {\it not} isometries of
the twistor space. HKC isometries that commute with the
homothety and the $k^3$ isometry do not depend on the coordinate $z$, and
{\em do} descend to isometries on the twistor space. A general analysis of
the HKC Killing equation leads to the following form for
HKC Killing vectors:
\be
k^i=-i\mu^i \ , \qquad k^z=i(K_i\mu^i-\mu)\ , \label{shk-isom}
\ee
where $\mu(u,\bar u)$ is a real function on ${\cal Z}$, with $\mu_i=\pa_i
\mu$ and $\mu^i=\mu_{\bar \jmath}K^{\bar \jmath i}$, satisfying
\be
D_i \pa_j \mu=0 \ . \label{Ddg=0}
\ee
Other HKC isometries depend explicitly on $z$ (it turns
out that they can be encoded in a holomorphic function and a
holomorphic one-form on the twistor space) and, with the exception of
the Sp(1) isometries, are disregarded in what follows (physically, they
cannot be gauged by coupling to an $N=2$ vector multiplet).
{From} \eqn{Ddg=0} one can prove that the vector \eqn{shk-isom} is
holomorphic. Furthermore it follows that the hyperk\"ahler potential
$\chi$ is invariant whereas $K(u,\bar u)$ changes by a \Ka\
transformation. Hence the twistor space $\cal Z$ admits an isometry
generated by
\be
\label{Z-isom}
{k}_i=i\,\pa_i \mu
\ee
and its complex conjugate. The Killing equation on $\cal Z$
can be verified directly from \eqn{Ddg=0}. We note that the moment
map of this isometry is the function $\mu$ itself. The case of constant
$\mu$ corresponds to the $k^3$ isometry, which acts trivially on $\cal Z$.

If, in addition, the isometry \eqn{shk-isom} is
triholomorphic in the hyperk\"ahler cone, then there is an extra
constraint on $\mu$:
\be
\label{Lk-X}
{\cal L}_{k}X_i\equiv -i \mu^j\pa_j X_i-i \pa_i \mu^j\,X_j= -2i
(K_j\mu^j-\mu)X_i\ ;
\ee
equivalently, ${\hat \omega}_{ij}\mu^j{\rm e}^{2K}+X_j\,D_i\,\mu^j
+2X_i\,\mu=0$. Triholomorphic HKC isometries thus
always descend to holomorphic isometries on ${\cal Z}$.

To end this section, we turn to the flat hyperk\"ahler cone whose
quantities of interest were defined in \eqn{pseudoRiemannian}, and
demonstrate explicitly that the corresponding twistor spaces are the
complex projective spaces $CP^{2n-1}$ (or their noncompact versions). We
start by singling out two of the complex coordinates with positive metric,
say $z^{2n}$ and $z^{2n-1}$, and bring the hyperk\"ahler potential and the
holomorphic two-form into the form
\bea
\chi(z,\bar z) &=& \sum_{i,j=1}^{2n-2} \eta_{i\bar\jmath} \,z^i \bar
z^{\bar \jmath}   + z^{2n-1}\bar z^{2n-1} + z^{2n} \bar z^{2n}\ ,
\nonumber \\
\Omega &=& \ft12 \omega_{ij} \, {\rm d} z^i \wedge {\rm d} z^j +
\Omega_{iz} \, {\rm d} z^i \wedge {\rm d} z^{2n} \equiv \sum_{i=1}^n {\rm
d} z^{2i-1} \wedge {\rm d} z^{2i}\ .
\eea
We now substitute
\be
z^{2n} = {\rm e}^z\ ,\qquad z^i =
{\rm e}^z\, u^i \ ,  \qquad (i= 1,\ldots,2n-1)
\ee
and find
\bea
\chi(z,\bar z, u,\bar u) &=& \exp [ z+ \bar z + K(u,\bar u)] \ ,
\nonumber \\
\Omega &=& {\rm e}^{2z} \Big[  \ft12 \omega_{ij} \, {\rm d}u^i \wedge {\rm
d}u^j - (\omega_{ij} u^j + \Omega_{iz})\,  {\rm d}z \wedge {\rm d}u^i
\Big] \ ,
\eea
where
\be
\label{K-CPn}
K(u,\bar u) = \ln\Big[ 1 + \sum_{i,j=1}^{2n-1}\eta_{i\bar \jmath} \,u^i
\,\bar u^{\bar \jmath}\Big]\ .
\ee

For $\eta_{i\bar\jmath}=\d_{i\bar\jmath}$, the \Ka\ potential $K(u,\bar
u)$ of the twistor space $\cal Z$ is the \Ka\ potential of $CP^{2n-1}$
(similarly for the indefinite case) and the metric is
\be
K_{i\bar \jmath} = {1 \over 1 +
\eta_{m\bar n}\, u^m\,\bar u^{\bar n} }
\left(\eta_{i\bar\jmath} - { \eta_{i\bar k}\,\bar u^{\bar k} \,
u^{l}\,\eta_{l\bar\jmath} \over  1 +
\eta_{p\bar q}\, u^p\,\bar u^{\bar q} }\right)\ .
\ee
The determinant of the metric is
\be
\det (K_{i\bar\jmath}) = \det (\eta_{i\bar\jmath})\, [1 +
\eta_{m\bar n}\, u^m\,\bar u^{\bar n}]^{-2n} = \det
(\eta_{i\bar\jmath})\, {\rm e}^{-2n \,K(u.\bar u)}\ ,
\ee
in accord with \eqn{K-equation}. The inverse metric is
\be
K^{{\bar \imath}j}=(1+\eta_{m \bar n} \,u^m \, \bar u^{\bar n}) \,
[\eta^{{\bar \imath}j}+{\bar u}^{\bar \imath} u^j]\ ,
\ee
where $\eta^{{\bar \imath}j}$ is the inverse of
$\eta_{i\bar\jmath}$. {From} the holomorphic two-form in these coordinates
one can read off the holomorphic one-form
\be
X=  (\omega_{ij} u^j + \Omega_{iz})\, {\rm
d}u^i\equiv \sum_{i=1}^{n-1}(u^{2i}{\rm d} u^{2i-1}-u^{2i-1}{\rm d} u^{2i})
+ \, {\rm d}u^{2n-1}\ .
\ee
Finally one can easily verify that $Y^i= \d^{i,2n-1}$, and that all
relations in \eqn{first-id} and \eqn{add-id} are satisfied.

\section{Quaternion-\Ka\ geometry}
\setcounter{equation}{0}

In this section, we construct the
quaternion-\Ka\ manifold $\Quat^{4(n-1)}$ and its geometry. We begin
by writing the horizontal metric and two-forms of the hyperk\"ahler cone
\cite{DWKV} in the special coordinates $(u^i,z)$ introduced in the previous
section. The metric $G_{a\bar b}$ \eqn{quaternionhorizontal} becomes
\be
\label{Q-metric}
G_{i{\bar \jmath}}=K_{i{\bar \jmath}}-{\rm e}^{-2K}X_i{X}_{\bar \jmath}\,,
\qquad  G_{z{\bar \imath}}=G_{z{\bar z}}=0 \ ,
\ee
which is manifestly orthogonal to the homothetic Killing vector
and the Sp(1) Killing vectors $k^3$ and $k^{\pm i}$, and hence
\be
\label{XGzero}
X^iG_{i{\bar \jmath}}=X^{\bar \imath} G_{i{\bar \jmath}} = 0\ .
\ee
Moreover, the horizontal metric $G$ is invariant under the Sp(1)
diffeomorphisms:
\be
\label{LX-G}
\left({\cal L}_{k^\pm} G\right)_{i\bar\jmath} = \left({\cal L}_{k^\pm}
G\right)_{ij}= 0\ .
\ee
Because $G_{i\bar\jmath}$ is $z$-independent and $G_{z\bar \imath}=0$,
the Lie derivative involves only the components  $k^{\pm\,i}$.
This identity follows from the equations
\eqn{Y-hat-omega},\eqn{first-id},\eqn{DX} and \eqn{XGzero}.

The horizontal two-forms $\vec{\cal Q}_{AB}$ \eqn{QK-cc} become
\be
{\cal Q}^3_{i\bar \jmath}=-i\,G_{i\bar \jmath}\ ,
\qquad {\cal Q}^3_{i{\bar z}}={\cal Q}^3_{z{\bar \imath}}=
{\cal Q}^+_{zi}=0\ , \qquad {\cal Q}^+_{ij}= {\rm e}^{z-\bar z}{\rm e}^K\,
\hat\omega_{ij}  \ ,
\label{almost-cs}
\ee
and satisfy
\be
X^i{\cal Q}^3_{i\bar \jmath}={\cal Q}^3_{i\bar \jmath}X^{\bar \jmath}=
X^i{\cal Q}^+_{ij}=0\ .
\ee
These two-forms rotate under Sp(1) according to
\be
\label{Sp1-rot}
{\cal L}_{k^-} {\cal Q}^+=-i{\cal Q}^3\ ,\qquad
{\cal L}_{k^+} {\cal Q}^3=-2i {\cal Q}^+\ .
\ee
Note that the $z$-dependence of ${\cal Q}^+_{ij}$ is relevant here, as $z$
transforms under the $k^-$ isometry.

The $\vec{\cal Q}$ are not covariantly constant in the twistor space
but satisfy the relations
\bea
D_k \Q^3_{i\bar\jmath}  &=&  i {\cal V}^-_{\bar\jmath}\,  \Q^+_{ik}\ ,
\nonumber \\
D_k \Q^+_{ij}  &=&  i {\cal V}^3_k\, \Q^+_{ij} \ ,  \nonumber\\
D_{k} \Q^-_{\bar\imath\bar\jmath} &=&-i  {\cal V}^3_k\,
\Q_{\bar\imath\bar\jmath}^- +2i   {\cal V}^-_{[\bar\imath}\,
\Q^3_{\bar\jmath] k} \ , \label{DQ-twistor}
\eea
where the Sp(1) connections $\vec{\cal V}$ were defined in \eqn{Sp1-conn} and
the covariant derivatives are defined with the twistor space affine
connection $\gamma_{ij}{}^k$. In \cite{DWKV} it was shown that there exists
another affine connection with respect to which the $\vec\Q$ are Sp(1)
covariantly constant. This connection projects to the affine connection of
$\Quat^{4(n-1)}$.

Now we project the metric and two-forms onto the quaternion-\Ka\ manifold
$\Quat^{4(n-1)}$. This space can be described as the subspace of the
twistor space
${\cal Z}$ orthogonal to the vector $X^i$.
As $X^i$ is neither holomorphic nor Killing, we cannot
perform a quotient\footnote{
           If the latter were a holomorphic vector field, a canonical way to
           project to the horizontal subspace would be to define a coordinate
           $\zeta$ by $X^i\pa_i\equiv \pa_\zeta$, in the same way as for the
           $k^3$ isometry (c.f. \eqn{def-z}). After performing the quotient,
           the metric $G_{i\bar \jmath}$ would then become horizontal with
           respect to the $\zeta$-direction and independent of $\zeta$ because
           of \eqn{LX-G}, and the  quaternion-\Ka\ space would be obtained
           directly.}. 
Fortunately, there is a holomorphic
vector field $Y^i(u)$, that we can use to single out a suitable
coordinate $\zeta$, and we define
\be
Y^i(v, \zeta) \,{\pa\over\pa u^i}\equiv {\pa\over \pa \zeta} \ .
\label{defzeta}
\ee
This choice of $\zeta$ is canonical but not unique.

In this way we decompose the holomorphic coordinates $u^i$ of the
Einstein-\Ka\ manifold into a special holomorphic coordinate $\zeta$
and $2n-2$ remaining ones $v^\a$, such that
the $(4n-4)$-dimensional manifold parametrized by the coordinates
$v^\a$ and $\bar v^{\bar \a}$ is quaternion-K\"ahler. However, we still
have to fix the dependence on the coordinates $z-\bar z$ and
$\zeta$ by choosing a suitable Sp(1) gauge condition.

In the new coordinates $(v^\a,\zeta)$ the vector
$Y^i= (0,\ldots,0,1)$. It then follows from the first equation of
\eqn{first-id} that $X_\zeta=1$. Using  \eqn{DX} and \eqn{first-id} we
then derive that
\be
\pa_\zeta X_i = Y^j \,\pa_jX_i = Y^j\,\pa_iX_j =0\ .
\ee
These results are summarized by ($\a= 1,\ldots,2n-2$),
\be
Y^\zeta=1\,,\quad Y^\a =0\,,\quad X_\zeta=1\,,\quad X_\a =
X_\a(v)\ . \label{canonical-Y}
\ee
The identities \eqn{add-id} now lead to five more (dependent)
relations. Defining
\be
Z^\a = \hat\omega^{\a\zeta}\ ,
\ee
these relations read as follows,
\bea
\label{Z-relations}
\omega_{\a\zeta}&=& 0\ ,\nonumber \\
\hat \omega^{\a\g}\,\omega_{\g\beta} &=& - \d_\b^\a\ ,\nonumber\\
Z^\a\,X_\a &=& 0\ , \nonumber \\
\omega_{\a\b} \, Z^{\b} &=& X_\a   \ ,\nonumber\\
\hat \omega^{\a\b} \,X_\b &=&- Z^{\a}\ .
\eea
Subsequently one proves that
\be
\pa_\zeta \omega_{\a\b} = -2\pa_{[\a}\omega_{\b]\zeta} = 0\ ,
\ee
so that $\omega_{\a\b}$ does not depend on $\zeta$.

The above equations
then show that $Z^\a$ and $\hat \omega^{\a\b}$ are also independent of
$\zeta$, so that we obtain the following decompositions for
the HKC  holomorphic tensors $\Omega_{ab}$ and $\Omega^{ab}$,
\bea
\Omega_{ab}(z,v)&=& {\rm e}^{2z}
\pmatrix{\omega_{\a\b}(v)&0&X_\a(v)\cr
\noalign{\vskip3mm}
0&0 &1\cr
\noalign{\vskip3mm}
-X_\b(v) &-1 & 0 \cr}\ , \nonumber\\[3mm]
\Omega^{ab}(z,v)&=& {\rm e}^{-2z}\pmatrix{ \hat\omega^{\a\b}(v)&
Z^\a(v)  & 0\cr
\noalign{\vskip3mm}
       -Z^\b(v)& 0&1\cr
\noalign{\vskip3mm}
0 &-1 & 0 \cr}\ ,
\eea
where
\be
\omega_{\a\b}(v)  = - \pa_{[\a}X_{\b]}(v)\ .
\ee
Observe that these tensors are thus entirely expressed in terms of the
$X_\a(v)$.

{For} later use, we note the following identities, which follow
from the second equation in \eqn{Y-hat-omega},
\bea
\label{X-upper}
X^\a&=& (\hat\omega^{\a\b} K_\b + Z^\a K_\zeta)\, {\rm e}^{2K}\ ,
\nonumber \\
X^\zeta &=& (1- Z^\a K_\a )\, {\rm e}^{2K}\ .
\eea

We now construct the quaternion-K\"ahler metric and two-forms
from the horizontal metric and two-forms by imposing
Sp(1) gauge conditions\footnote{As the Sp(1) isometry of the
hyperk\"ahler cone is lost when descending to the twistor space $\Z$,
one may wonder how a {\it gauge} symmetry corresponding to $k^\pm$
can act on the twistor space. Mathematically, this
happens because the $S^2 \equiv $ Sp(1)/U(1) bundle is non-trivial.
Physically, if we gauge the full Sp(1) in the hyperk\"ahler cone,
after eliminating the U(1) connection (to descend to $\Z$),
the remaining connections corresponding to the
coset generators $k^\pm$ have gauge transformations that include a term
proportional to the U(1) connection. Since this connection is determined
in terms of the coordinates of $\Z$, we can compute its curvature; we find
the K\"ahler form of $\Z$. Consequently, this inhomogeneous term in the
transformations of the coset connections is an obstruction to finding a
rigid isometry, but clearly we can still choose a {\it local} gauge fixing
condition for the coset generators.}. A convenient set of conditions
is $z-\bar z =\zeta=0$. This gauge can indeed be chosen by using the
remaining $k^{-i}\propto X^i$ symmetry, since $X^\zeta $ is generically
non-vanishing. As discussed above, because $\zeta$ is not a coordinate
along a Killing vector, the metric $G_{i\bar \jmath}$ is in general
not independent of $\zeta$ and $G_{\zeta \bar \imath}$
is non-zero. Hence, the quaternion-\Ka\ metric is obtained by
setting $\zeta=0 \Rightarrow {\rm d}\zeta =0$. Since this is a
holomorphic gauge choice, the metric is still hermitean, and
its components are given by
\be
\label{QK-metric}
G_{\alpha {\bar \beta}}=K_{\alpha {\bar \beta}}-{\rm e}^{-2K}X_\alpha
X_{\bar \beta}\ .
\ee
This metric is non-degenerate; its inverse can be expressed
in terms of $K^{{\bar \a}\beta}=(K_{\a{\bar \b}})^{-1}$,
\be
G^{{\bar \a}\b}=K^{{\bar \a}\b}+{\rm e}^{-2K}\frac{K^{{\bar \a}\g}X_\g
K^{{\bar \d}\b}X_{\bar \d}}{1-{\rm e}^{-2K}X_{\bar \d}K^{{\bar \d}\g}X_\g}\ .
\ee
The quaternion-\Ka\ two-forms are given by
\be \label{QK-forms}
{\cal Q}^3_{\a\bar \b}=-iG_{\a\bar \b}\qquad
{\cal Q}^+_{\a\b}={\rm e}^{-K}(\o_{\a\b}+2K_{[\a}X_{\b]})\ .
\ee
We suppress the $+$ superscript on the tensor ${\cal Q}_{\a\b}^+$ below,
as its holomorphic indices indicate that we are dealing with
$\Q^+$. The two-form ${\cal Q}$ should be
non-degenerate, such that there exist an inverse ${\cal Q}^{\a\b}$,
\be
{\cal Q}^{\a \g}{\cal Q}_{\g \b}=-\delta^\a{}_\b\ .
\ee
This inverse tensor can be found explicitly. Using
\eqn{Z-relations} and \eqn{X-upper}, one can verify that it takes the
form
\be
{\cal Q}^{\a \b}={\rm e}^K\left[ {\hat \o}^{\a\b}-2\frac{Z^{[\a}X^{\b]}}
{X^\z}\right]\ ,
\ee
and that it is related to ${\cal Q}^-\equiv {\bar \Q}$ by
\be
{\cal Q}^{\a\b}={\bar {\cal Q}}_{\bar \g \bar \d}\,G^{\bar \g \a}
\, G^{\bar \d \b}\ .
\ee
This property ensures that the quaternionic algebra holds. Note that the
expression for the inverse is non-degenerate when $X^\zeta$ is nonvanishing.
According to \eqn{X-upper}, this is so when $Z^\a K_\a \neq 1$, which is
generically the case because $Z^\a$ is holomorphic and $K_\a$ is not.

The Sp(1) connections of the quaternion-\Ka\ space follow
from \eqn{Sp1-conn},
\be
{\cal V}_\a^3=- iK_{\a}\,,\qquad
{\cal V}_\a^+=  {\rm e}^{-K}\,X_\a\ ,
\ee
and their complex conjugates.
The Sp(1) curvature two-forms are then defined in the
quaternion-\Ka\ space by
\be
{\cal R}^3 \equiv {\rm d}{\cal V}^3-2i{\cal V}^+\wedge {\cal V}^-\ ,
\qquad
{\cal R}^+ \equiv {\rm d}{\cal V}^+-i{\cal V}^3\wedge {\cal V}^+\ .
\ee
and satisfy
\be
{\vec {\cal Q}}=-\ft12 {\vec {\cal R}}\ .
\ee
These formulae can be derived both in the twistor
and quaternion-\Ka\ spaces because the Bianchi identities,
\bea
{\rm d}{\cal R}^3 = 2i({\cal V}^-\wedge {\cal R}^++{\cal V}^+
\wedge {\cal R}^-)\ , \qquad
{\rm d}{\cal R}^+ = i({\cal V}^+\wedge {\cal R}^3+{\cal V}^3
\wedge {\cal R}^+)\ ,
\eea
hold in both cases (see \eqn{DQ-twistor}).

In section 4 we have shown that triholomorphic HKC isometries descend to
holomorphic isometries on the twistor space ${\cal Z}$, with the additional
constraint \eqn{Lk-X}. We now study how these isometries descend to the
quaternion-\Ka\ manifold and how they give rise to quaternionic isometries
({\it i.e.}, isometries that leave the quaternionic structure invariant up
to an Sp(1) rotation). The fact that the triholomorphic HKC isometries lead
to quaternionic isometries is known from the mathematics literature
\cite{Swann}, and can be understood from the superconformal calculus
\cite{DWKV}.

We start by observing that $G_{i\bar \jmath}$ is preserved by the
triholomorphic isometry \eqn{Z-isom}. Indeed, using \eqn{Lk-X}, we find that
\be
\label{Lk-G}
{\cal L}_{k} G_{i\bar \jmath}=0\ .
\ee
Here we take the Lie derivative along the total vector field $k$
comprising both $k^i$ and $k^{\bar \imath}$. Obviously, the action of
this isometry is not in general restricted to the
quaternion-\Ka\ subspace, because the coordinate $\zeta$, which
has been put to zero by the gauge choice, may change. To
correct for this we have to add a {\it compensating} Sp(1)
transformation (with a coordinate-dependent coefficient) to restore
the $\zeta=0$ gauge. Because the Sp(1) transformation takes the form
$\d u^i\propto X^i$, we must thus combine the action of the isometry
associated to $k_i=i\partial_i \mu\equiv i\mu_i$ (see \eqn{Z-isom}) with
the following compensating Sp(1) transformation:
\be
\delta \zeta=-k^\zeta =i\mu^\zeta \ ,\qquad
\delta u^\a=-\frac{k^\z}{X^\z}X^\a=i\frac{\mu^\zeta}{X^\z}X^\a\ .
\label{comp-sp-1}
\ee
This modification leaves $G_{i\bar \jmath}$ invariant because
$X^iG_{i\bar \jmath}=0$ and hence the Lie derivative of the metric
$G_{i\bar\jmath}$ along $f X^i$ for {\it any} function $f$ of the coordinates
(see \eqn{LX-G}) vanishes. Thus we conclude that the vector defined by
\be
\label{Q-isom}
{\hat k}^\a=k^\a-\frac{k^\z}{X^\z}X^\a=-i \Big[ \mu^\a - \mu^\z
\,{X^\a\over X^\z}\Big] \ ,\qquad \hat k^\zeta=0\ ,
\ee
preserves the gauge $\zeta=0$, and
therefore defines an isometry of the quaternion-\Ka\ metric:
\be
{\cal L}_{\hat k}G_{\a\bar \b}=0\ .
\ee

A similar result can be derived for the action of the isometries on
the quaternionic structure. Following the same procedure as for the metric,
we first determine the variation with respect to the isometry \eqn{Z-isom}
of ${\cal Q}^3_{i\bar \jmath}$ and ${\cal Q}^+_{ij}$ (see \eqn{almost-cs}),
and find (provided we also take the variations of $z$ and $\bar z$ into
account) that they are invariant:
\be
{\cal L}_{k}\vec{\cal Q} =0\ .
\ee
However, these isometries do not preserve the gauge $z-\bar z =\zeta =
0$. Hence we must introduce an infinitesimal compensating Sp(1)
transformation to restore the gauge conditions, one as in \eqn{comp-sp-1}
for $\zeta=0$ and a similar one for $z-\bar z=0$. The
combined effect of the projected HKC  isometry and the
compensating Sp(1) transformation rotates the $\vec \Q$
by an Sp(1) rotation. Restricting the twistor space forms $\vec \Q$ to the
quaternion-\Ka\ ones (see \eqn{QK-forms}), we thus derive the following
result for the quaternionic structure,
\bea
\label{rot-twoforms}
{\cal L}_{\hat k}{\cal Q}^3&=&  -2 \,{\rm e}^K \Big[ {\mu^{\bar \z}\over
\bar X^{\bar \z}} \,{\cal Q}^+ + {\mu^\z\over X^\z}\,
{\cal Q}^-\Big]  \ ,\\
{\cal L}_{\hat k}{\cal Q}^+&=&-i\Big[K_i(\mu^i- \mu^\z {X^i\over X^\z})
+ K_{\bar \imath}(\mu^{\bar \imath}- \mu^{\bar \z}{X^{\bar
\imath} \over X^{\bar \z}})-2\mu\Big]\,{\cal Q}^+ +\, {{\rm e}^K
\,\mu^\z\over X^\z} \, {\cal Q}^3 \ . \nonumber
\eea

To end this section it is instructive to return to the example
based on a  hyperk\"ahler cone with a flat $(4n)$-dimensional
(pseudo)-Riemannian metric. The twistor
space $\cal Z$ associated with the underlying quaternion-\Ka\
manifold is a (noncompact) $CP^{2n-1}$ and was discussed in the
previous section. Now we construct the underlying
quaternion-\Ka\ manifold. At the end of the previous section, we found
that the coordinate $\zeta=u^{2n-1}$. It is now straightforward
to determine the metric from \eqn{QK-metric} by restricting the
coordinates to the remaining ones, denoted by $u^\a$ with $\a=
1,\ldots ,2n-2$, and putting $\zeta=0$. The result is
\be
G_{\a\bar \b} = {\eta_{\a\bar\b} \over 1 + \eta_{\g\bar\d} \,
v^\g\,\bar v^{\bar\d} }
-{ \eta_{\a\bar\g} \bar v^{\bar\g} \;
v^{\d}  \eta_{\d\bar \b}   - \omega_{\a\g} v^\g \;
\bar v^{\bar\d}  \bar \omega_{\bar\d\bar\b}
\over (1 + \eta_{\g\bar\d} \, v^\g\,\bar v^{\bar\d})^2 }   \ .
\ee
This is the metric for (noncompact) $HP(n-1)$.
The Sp(1) curvature can also easily be computed,
\be
{\cal R}^+=- \left[ {\o_{\a\b}\over 1 + \eta_{\g\bar\d} \,
v^\g\,\bar v^{\bar\d} } - 2\,  {\eta_{\a\bar \gamma}
{\bar v}^{\bar \gamma}\; v^\delta \o_{\delta\beta} \over (1 +
\eta_{\g\bar\d} \,  v^\g\,\bar v^{\bar\d})^2 }\right] \;
{\rm d}v^\alpha\wedge {\rm d}v^\beta\ .
\ee

A particular case of this series is the four-sphere $S^4=HP^1$, so
$n=2$. As this space is compact, we take $\eta_{\a{\bar
\b}}=\delta_{\a{\bar \b}}$ and $\o_{12}=1$. The metric
simplifies to
\be
{\rm d}s^2=\frac{{\rm d}v^\alpha\,{\rm d}{\bar v}^{\bar \alpha}}
{(1+|v|^2)^2}\ ,
\ee
which is indeed the conformally flat metric on $S^4$. The Sp(1) curvature
takes the simple form
\be
{\cal R}^3=2i\frac{{\rm d}v^\alpha\wedge {\rm d}{\bar v}^{\bar \alpha}}
{(1+|v|^2)^2}\ ,\qquad
{\cal R}^+=-2\,\frac{{\rm d}v^1\wedge {\rm d}v^2}
{(1+|v|^2)^2}\ .
\ee

For the noncompact case, namely four-dimensional anti-de-Sitter space,
we choose $\eta_{\alpha \bar \beta}=-\delta_{\alpha \bar \beta}$.
The metric components then are, with $v^1=u$ and $v^2=v$,
\be
G_{u\bar u}=G_{v\bar v}=-\frac{1}{1-u\bar u-v\bar v}\ ,
\qquad G_{u\bar v}=0\ .
\ee

\section{Superconformal tensor multiplets}
\setcounter{equation}{0}
Metrics on $4n$ (real) dimensional hyperk\"ahler cones with $n$
commuting triholomorphic isometries can be constructed (locally)
by a duality transformation of a general superconformally invariant $N=2$
tensor multiplet action. Every hyperk\"ahler cone with such isometries can
be obtained in this way.

An $N=2$ tensor supermultiplet consists of three scalar fields, which we
group into a real scalar $x$ and a complex one $v$, and a tensor gauge field
$B_{\m\n}$ with corresponding gauge-invariant field strength $H^\m =
-\ft12 \varepsilon^{\m\n\rho\sigma}\,\pa_\n B_{\rho\sigma}$. Furthermore it
contains a doublet of Majorana spinors and its (minimal) off-shell version
requires an auxiliary complex scalar field. The first part of the discussion
below closely follows \cite{F2,Hitchin}.

To obtain a hyperk\"ahler cone one starts from a system of
hypermultiplets that is invariant under (rigid) superconformal
transformations. However, for the moment we ignore the superconformal
aspects and proceed to describe the most general couplings of tensor
multiplets. The restrictions imposed by superconformal invariance are
introduced below. In $N=1$ superspace, a general Lagrangian for tensor
multiplets is a real function $F(x^I, v^I, \bar v^I)$, where indices
$I,J,\ldots$ label the $n$ tensor multiplets, such that $F$ satisfies the
following differential equation:
\be
F_{x^Ix^J}+F_{v^I{\bar v}^J}=0\ .
\label{laplace}
\ee
Here the subscripts denote differentiation with respect to the
corresponding fields. Observe that this constraint implies that the mixed
derivative with respect to $v^I$ and $\bar v^J$ is symmetric in the indices
$I$ and $J$.

The bosonic part of the Lagrangian for the $n$ tensor supermultiplets
is
\be
{\cal L} = F_{x^Ix^J}\Big(\pa_\m v^I \pa^\m {\bar v}^I+\ft14(\pa_\m x^I
\pa^\m x^J -H_\m ^IH^{\m J})\Big)
+\ft{1}{2}i\Big(F_{v^Ix^J}\pa_\m v^I-F_{{\bar v}^Ix^J}\pa_\m {\bar
v}^I\Big)H^{\m J}\ .
\label{tensor-L}
\ee
For fixed $x^I$, the corresponding nonlinear sigma model is a \Ka\
space with \Ka\ potential equal to $F(v^I,\bar v^I)$, whereas for
fixed $v$, one gets the bosonic part of the supersymmetric Lagrangian
for $N=1$ tensor multiplets. After adding a total derivative,
the term linear in $H^{\m I}$ can be rewritten as
\be
{\cal L}^\prime
=\ft{1}{4}i \Big(F_{v^Ix^Jx^K}\pa_\m v^I\pa_\n x^K -F_{{\bar
v}^Ix^Jx^K}\pa_\m {\bar v}^I\pa_\n x^K - 2 F_{x^Ix^Jx^K} \pa_{\m} \bar
v^I \pa_{\n} v^K \Big)
\varepsilon^{\m\n\rho\sigma}\, B_{\rho\sigma}^J \ .
\label{F-B}
\ee

We now dualize the vectors $H^{\mu I}$ by introducing real multipliers
$y_I$, and adding to the action a term $-\ft12 y_I\pa_\m H^{\m
I}$. The field equations for the $y_I$ now ensure that the quantities
$H^{\m I}$ satisfy the constraint $\pa_\m H^{\m I}=0$. However, rather
then imposing the field equations for the $y_I$ one can solve the
field equations for the $H^{\m I}$. This yields the Lagrangian,
\bea
{\cal L}&=& F_{x^Ix^J}\Big(\pa_\m v^I \pa^\m {\bar v}^J+\ft14\pa_\m
x^I \pa^\m x^J \Big)\nonumber \\
&&\ +\ \ft14 F^{x^Ix^J}\Big(\pa_\m
y_I+i(\pa_\m v^K F_{v^Kx^I} - F_{x^I{\bar v}^K}\pa_\m {\bar v}^K)
\Big) \nonumber \\
&& \hspace{16mm} \times \Big(\pa^\m y_J+i(\pa^\m
v^L F_{v^Lx^J} -F_{x^J{\bar v}^L}\pa^\m {\bar v}^L) \Big)\ ,
\eea
where $F^{x^Ix^J}$ denotes the inverse of $F_{x^Ix^J}$.

At this point we introduce a second set of $n$ complex fields $w_I$ by
the conditions
\be w_I = \ft12 \Big(i y_I + \frac{\pa F}{\pa x^I}
\Big) \ ,
\label{def-w}
\ee
which determine $x^I$ and $y_I$ in terms of the $v^I$, $w_I$,
$\bar v^I$ and $\bar w_I$. The definition \eqn{def-w} may seem
somewhat arbitrary but is obvious when performing the duality
transformation in terms of $N=1$ superfields. Note that the
metric describing the target space of the nonlinear sigma model with
coordinates $v^I$ and $w_I$ is independent of $w_I- \bar w_I$, and
hence the space has $n$ commuting isometries associated with shifts of
$w_I$ by imaginary constants. Moreover, as the action is $N=2$
supersymmetric, the space is hyperk\"ahler.

To evaluate the metric in terms of the complex coordinates, we vary
\eqn{def-w} and find \be \delta x^I = F^{x^Ix^J}\left( (\delta w_J +
\delta\bar w_J) - \delta v^K F_{v^Kx^J}-F_{x^J{\bar v}^K} \delta \bar
v^K\right) \ .  \ee With the help of this equation the metric follows
straightforwardly:
\bea
g_{v^I{\bar v}^J}&=&-\Big(F_{x^Ix^J}+F_{v^Ix^K}F^{x^Kx^L} F_{x^L{\bar
v}^J}\Big)\,\nonumber\\ g_{v^I{\bar w}_J}&=&F_{v^Ix^K}F^{x^Kx^J}
\ ,\nonumber\\ g_{w_I{\bar v}^J} &=& F^{x^Ix^K}F_{x^K{\bar v}^J} \ ,
\nonumber\\ g_{w_I{\bar w}_J}&=&-F^{x^Ix^J}\ .
\eea
The inverse metric can be computed directly and its components are
\begin{eqnarray}
g^{{\bar v}^Iv^J}&=&-F^{x^Ix^J}\ , \nonumber\\ g^{{\bar
v}^Iw_J}&=&-F^{x^Ix^K}F_{x^Kv^J}\ , \nonumber\\ g^{{\bar
w}_Iv^J}&=&-F_{{\bar v}^Ix^K}F^{x^Kx^J}\ , \nonumber\\ g^{{\bar
w}_Iw_J}&=&-(F_{x^Ix^J}+F_{{\bar v}^Ix^K}F^{x^Kx^L}F_{x^Lv^J})\ .
\end{eqnarray}
Finally one easily verifies that this space is a \Ka\ space with \Ka\
potential equal to \be K(v,w, \bar v, \bar w)= F(x,v,\bar v) -
(w_I+{\bar w}_I)\, x^I\ .
\label{dual-KP}
\ee For further discussion of the geometric properties of this
hyperk\"ahler space we refer to \cite{Hitchin}.

We now discuss the superconformal couplings of the $n$
tensor multiplets. The scaling weight of the tensor gauge field is
fixed so that it is compatible with its gauge invariance, and hence
the field strengths scale with weight $3$. To have scale invariance
for the full action the scalar fields $x^I$ and $v^I$ must therefore
have weight $2$, whereas the second derivatives of the function $F$
should scale with weight $-2$ such that the Lagrangian scales
uniformly with weight $4$ (recall a spacetime derivative has weight
$+1$). Therefore it follows that the derivatives $F_{x^I x^J}$ and
$F_{x^I v^J}$ must be homogeneous functions of $x^I$, $v^I$ and $\bar
v^I$ of degree $-1$. This condition is not yet
sufficient for superconformal invariance, because we know from $N=1$
superconformal symmetry that the theory must also be invariant under
phase transformations of the complex fields $v^I$. Hence
$N=2$ superconformal invariance requires both \eqn{laplace} and
the identities \eqn{homo-so2} presented in appendix~A; these
are implied by the invariance and homogeneity requirements discussed above.

Subsequently, in appendix A, we analyze the identities \eqn{homo-so2}
and show that, modulo irrelevant terms in $F$ that do {\it not} contribute to
the action, they imply that the function $F(x,v,\bar v)$ is homogeneous of
first degree and invariant under phase transformations acting on the $v^I$.
Furthermore the irrelevant terms can be chosen to make
$F_{x^Iv^J}$ symmetric in $I$ and $J$. Henceforth we
restrict the function $F(x,v,\bar v)$ accordingly, which implies
that it satisfies the following differential equations:
\bea
F_{x^Ix^J} + F_{v^I\bar v^J} &=& 0\ , \nonumber \\
x^I F_{x^I} +  v^I F_{v^I} + \bar v^I F_{\bar v^I} &=& F
\ ,\nonumber\\
v^I F_{v^I} - \bar v^I F_{\bar v^I} &=& 0 \ , \nonumber\\
F_{x^Iv^J}-F_{x^Jv^I}&=&0\ .\label{inv-F}
\eea
Moreover, it turns out that the phase
transformations are part of an SU(2) group, under which the Lagrangian
is also invariant. Under the SU(2)
transformations $x^I$, $v^I$ and ${\bar v}^I$ transform as vectors,
\be
\d v^I = i\epsilon^3 v^I +
\epsilon^+x^I\,,\quad \d \bar v^I = -i\epsilon^3 \bar v^I +
\epsilon^- x^I\,,\quad  \d x^I = -2(\epsilon^-v^I + \epsilon^+\bar
v^I) \ ,
\ee
and leave $x^I x^J + 2 v^I\bar v^J + 2 \bar v^I v^J$ invariant for all $I$
and $J$. Equations \eqn{inv-F} imply that the Lagrangian \eqn{tensor-L}
is SU(2) invariant up to a total derivative. The only nontrivial
part of the calculation arises in proving the invariance
(modulo total derivatives) of the interaction linear in the tensor
field; an intermediate step is:
\be
\d \Big(F_{x^Iv^J}\pa_\m v^J - F_{x^I \bar v^J}\pa_\m \bar v^J  \Big)
= \epsilon^+\,\pa_\m F_{v^I}  - \mbox{ h.c. }\ .
\ee
We stress that this SU(2) invariance, which is crucial for $N=2$
superconformal invariance, is automatic at this point. Hence we
conclude that the Lagrangians based on functions satisfying
\eqn{inv-F} encode all the  $N=2$ superconformal theories of $n$
tensor supermultiplets.

When performing the duality transformation, SU(2) variations
proportional to the divergence of $H_\mu^I$ no longer vanish
identically and must be cancelled by assigning a suitable SU(2)
transformation to the fields $y_I$. This determines the SU(2)
transformations of the fields $v^I$, $w_I$ and their complex
conjugates. The U(1) transformations with parameter $\epsilon^3$
still act only on the $v^I$ by a uniform change of phase. Under the
remaining transformations we have
\be
\d v^I = \epsilon^+k^{-v^I}\,, \qquad  \d w_I = \epsilon^+
\,k^{-w_I} \ ,
\ee
with
\be
k^{-v^I} =  x^I\,,\qquad k^{-w_I} = F_{v^I} \ .
\ee
One can show, always using \eqn{inv-F}, that these transformation
rules correctly generate the SU(2) algebra. With lower indices, the
vectors take a very simple
form,
\be
k^+_{v^I} = 0 \,,\qquad k^+_{w_I} = 2\,v^I\ ,  \label{killing+}
\ee
where we made use of \eqn{inv-F}.  Observe
that this vector depends exclusively on the coordinates $v^I$
and is thus holomorphic. Because $k^{-v^I}$ and $k^{-w_I}$ do not depend
on the imaginary part of $w_I$, the SU(2) transformations commute with
the isometries associated with purely imaginary shifts of the
$w_I$. The scale transformations also commute with these isometries,
because $w_I$ has zero scaling weight.

We have thus shown that the resulting hyperk\"ahler manifold has an SU(2)
isometry group. Conversely one can show directly that the superconformal
invariance of the hypermultiplet theories is carried over to the tensor
multiplets. This is not the general situation with regard to other
invariances: the imaginary shifts of the fields $w_I$ in the hypermultiplet
description act trivially on the tensor multiplets, and symmetries that do
not commute with these shifts do not induce symmetries of the tensor
multiplet action.  This is similar to the situation that one has when
dualizing vector multiplets in three spacetime dimensions to hypermultiplets
\cite{Swa}.

We now show that the \Ka\ potential \eqn{dual-KP} is the
hyperk\"ahler potential of a hyperk\"ahler cone; this potential can be
rewritten as
\be
\chi(v,w, \bar v, \bar w)= F(x,v,\bar v) - (w_I+{\bar w}_I)\, x^I =
2v^I F_{v^I}(x,v,\bar v) \ ,
\label{hyper-KP}
\ee
and is invariant under SU(2) transformations, as one can explicitly check.
Under scale transformations the coordinates $v^I$ transform
with weight 2, whereas the $w_I$ are invariant, so that $\chi$ has weight two.
To verify the homothety equation \eqn{homothety}, we determine the homothetic
Killing vectors $(\chi^{v^I},\chi^{w_I})$ obtained from $\chi$ and establish
that they are holomorphic. The derivatives of the hyperk\"ahler potential are
\be
\label{chi-lower}
\chi_{v^I}= F_{v^I}\ , \qquad  \chi_{w_I}=-x^I\ .
\ee
Using \eqn{inv-F}, we find
\be
\label{chi-upper}
\chi^{v^I} =  2 v^I\ , \qquad  \chi^{w_I} = 0\ ,
\ee
which indeed are holomorphic. Furthermore,
we have the correct normalization conditions:
\be
\chi^{v^I} \chi_{v^I}+\chi^{w_I} \chi_{w_I} = k^{-v^I}k^+_{v^I} +
k^{-w_I}k^+_{w_I} = \chi \ ,
\ee
confirming that $\chi$ is the hyperk\"ahler potential.
Finally we can read off the complex structures, by using the relation
\eqn{su(2)-vector}. One finds that $J^3$ is indeed the canonical
one, {\it i.e.}, $J^{3a}{}_b=i\delta^a{}_b$, where now the indices $a,b$ run
over both $v^I$ and $w_I$. The holomorphic two-form, associated with $J^+$ is
\be
\label{tensor-twoform}
\O={\rm d}w_I\wedge {\rm d}v^I\ ,
\ee
where we made use of \eqn{killing+}.
Evidently, imaginary shifts of $w_I$ preserve the
complex structures, so that these isometries are
triholomorphic.

To illustrate the above results we present the simple example of a single
tensor multiplet \cite{philippe}. The corresponding
hyperk\"ahler cone is flat\footnote{
           A flat space can also be obtained from a tensor Lagrangian which is
           not conformally invariant. Namely, choose $F(x,v,\bar v) = v\bar v
           -\ft12 x^2$, which leads to $K(v,w,\bar v, \bar w) = v\bar v + w\bar
           w$, up to a \Ka\ transformation. However, the isometries associated
           with imaginary shifts of $w$ do not commute with SU(2) in this
           case.}.   
The example is based on the following homogeneous function,
invariant under phase transformations \cite{F2},
\be
F(x,v,\bar v) = r-x[ \ln(x +r) -\ft12 \ln(4\,v{\bar v})]\ ,
\label{F-HPn}
\ee
where $r={\sqrt {x^2+4\,v{\bar v}}}$. Its relevant derivatives are equal
to
\bea
&&
F_x = -\ln(x+r)+\ft12\ln(4\, v{\bar v})\ ,\qquad F_v= {r\over 2\,v}\ ,
\nonumber \\
&&
F_{xx} = -F_{v\bar v} = - {1\over r}\,,\qquad F_{xv} = {x\over 2\, v
r}  \ . \label{F''}
\eea
It is now easy to see that the Lagrangian \eqn{tensor-L} is invariant
under SU(2) transformations, with the exception of the term linear in
the tensor field strength. However, rewriting this term in the
form \eqn{F-B}, SU(2) invariance becomes manifest \cite{philippe}:
\bea
{\cal L}&=&- \frac1r \Big(\pa_\m v\, \pa^\m {\bar v}+\ft14(\pa_\m
x\, \pa^\m x - H_\m H^{\m })\Big)\nonumber\\
&& + \frac1{2r^3}\Big(\bar v \,\pa_\m v\,\pa_\n x + v\, \pa_\m x \,\pa_\n
{\bar v}  + x \, \pa_{\m} \bar v\, \pa_{\n} v\Big)  \,
i\, \varepsilon^{\m\n\rho\sigma}\, B_{\rho\sigma}   \ .
\eea

After the duality transformation we obtain the new variable $w$ whose
real part is given by $w+{\bar w}= F_x$.
Solving $x$ in terms of $w,{\bar w},v,{\bar v}$ gives
\be
x=- 2 {\sqrt {v{\bar v}}}\, \sinh (w+{\bar w})\ .
\ee
The hyperk\"ahler potential for the corresponding hyperk\"ahler cone
is equal to
\be
\chi(v,w,\bar v,\bar w) =F-(w+{\bar w})x =  2 \sqrt{v{\bar v}}\, \cosh
(w+{\bar w})\ .
\ee
Computing the line element, we find a flat metric
\bea
{\rm d}s^2&=&{\cosh(w+\bar w)\over \sqrt{v\bar v}}  \Big[ \ft12 {\rm d}v
\,{\rm d}\bar v
+ 2\,{v\bar v} \, {\rm d}w\,{\rm d}\bar w \Big]
+ {\sinh(w+\bar w)\over \sqrt{v\bar v}} \Big[ \bar v\, {\rm d}v\,
{\rm d}\bar w
+  v\, {\rm d}w\, {\rm d}\bar v \Big] \nonumber \\[2mm]
& =& \Big\vert {\rm d}({\rm e}^w \sqrt v)\Big\vert{}^2 + \Big\vert {\rm d}
({\rm e}^{-w}  \sqrt v)\Big\vert{}^2\ .
\eea

To end this section we turn to the question of finding explicit
realizations of the function $F$. As was shown in
\cite{KLR,Hitchin}, a function $F$ that satisfies \eqn{laplace} can be
represented as a contour integral:
\be
F={\rm Re}\oint\,\frac{{\rm d}\xi}{2\pi i \xi}\,G(\eta(\xi),\xi)\ ,
\label{contour-F}
\ee
where
\be
\eta^I(\xi)= {\bar v}^I\, \xi^{-1}  + x^I -  v^I\,\xi\ .
\ee
It follows by straightforward calculation that this form of $F$
satisfies \eqn{laplace} and that $F_{x^Iv^J}$ is symmetric in $I$
and $J$. Furthermore, the conformal constraints \eqn{inv-F} translate
into simple constraints on the function $G(\eta(\xi),\xi)$.
Firstly, the homogeneity constraint in \eqn{inv-F} requires $G$ to be
homogeneous of first degree in the variables $\eta$ (under the contour
integral),
\be
\oint\, \frac{{\rm d}\xi}{2\pi i \xi}\, \eta^I\frac{\partial G}
{\partial \eta^I}=\oint\,\frac{{\rm d}\xi}{2\pi i \xi}\, G\ .
\ee
Secondly, the SO(2) invariance of $F$ implies that $G(\eta(\xi),\xi)$ is
only a function of $\eta(\xi)$, so there is no explicit $\xi$ dependence.
Indeed, one computes
\bea
0 \ =\ ({\bar v}^I\pa_{{\bar v}^I}-v^I\pa_{v^I})F&=& \oint\,\frac{{\rm d}\xi}
{2\pi i\xi}\frac{\pa G}{\pa \eta^I} \Big({{\bar v}^I\over \xi}
+\xi\,v^I\Big)\nonumber\\[2mm]
&=&-\oint\,\frac{{\rm d}\xi}{2\pi i\xi}\,\frac{\pa G}{\pa \eta^I}
\,\xi\,\frac{\pa \eta^I}{\pa \xi}\nonumber\\[2mm]
&=&-\oint\,\frac{{\rm d}\xi}{2\pi i}\Big(\frac{{\rm d}}{{\rm d}\xi}G-
\frac{\pa G}{\pa \xi}\Big)\ .
\eea
The first term in the last line between brackets is zero because it is a
total derivative and the contour is closed. The whole expression therefore
vanishes if $G$ is only a function of $\eta(\xi)$ and there is no
explicit $\xi$ dependence\footnote{
           In fact, $\pa G/\pa \xi$ should only vanish under the contour
           integral, but in most interesting examples, we have the stronger
           relation $\pa G/\pa \xi=0$.}. 
For the hyperk\"ahler potential one thus obtains the expression
\be
\chi(v,\bar v, w, \bar w) = {\rm Re} \, \left[ -2 v^I \oint
\frac{{\rm d}\xi}{2\pi i} \; {\pa G(\eta) \over \pa \eta^I} \right] \,,
\ee
where the coordinates $w_I$ satisfy
\be
(w+\bar w)_I = {\rm Re}\oint\,\frac{{\rm d}\xi}{2\pi i \xi}\; {\pa
G(\eta) \over \pa \eta^I}  \ .
\ee

There are plenty of examples, the simplest one, corresponding to
\eqn{F-HPn}, is \cite{KLR}
\be
G(\eta) =\eta  \ln \eta\ , \label{H-eta}
\ee
with a contour turning clockwise and anticlockwise around the two
roots of $\eta(\xi)=0$ (contributions from a contour around the origin
are irrelevant as they lead to total derivatives in the action). This
function is indeed homogeneous under the contour integral since
terms linear in $\eta$ do not contribute to $F$.

The duality between tensor multiplets and hypermultiplets with
triholomorphic isometries bears a close relation to the ${\bf c}$-map,
which maps $N=2$ supersymmetric abelian vector multiplets to hypermultiplets,
again with triholomorphic isometries. The combined map\footnote{This
vector-tensor multiplet duality is actually more direct than either the
${\bf c}$-map or the tensor-hypermultiplet duality as it does not involve the
equations of motion.}
interchanges vector multiplets and tensor multiplets. The vector
multiplet Lagrangian is encoded in  a holomorphic function ${\cal F}(v)$,
where the complex vector multiplet scalars are denoted by $v^I$. The
corresponding function $F$ that characterizes the Lagrangian of the tensor
multiplets is given by
\be
F(x,v,\bar v) = {\rm Re}  \Big[- i\, \bar v^I\, {\cal F}_I(v) +
\ft12 i\, x^I x^J\, {\cal F}_{IJ}(v) \Big] \ ,
\ee
which obviously satisfies \eqn{laplace}. There is a corresponding function
$G$ (see \eqn{contour-F}), evaluated with a contour around the origin:
\be
G=\frac{i{\cal F}(\xi\eta)}{2\xi^2}\ .
\ee
Note that conformal
invariance is not preserved by this map. It is well-known that
the functions ${\cal F}(v)$ belong to certain equivalence classes
(via electric-magnetic duality); likewise the functions $G(\eta)$
fall into certain equivalence classes. Dualizing to the
corresponding hyperk\"ahler space in both cases, we find that these
equivalences are the same and arise from considering different sets of
triholomorphic isometries on the hyperk\"ahler space (see section 7).

\section{Abelian quaternionic isometries}
\setcounter{equation}{0}
In this section we discuss the quaternion-\Ka\ geometry associated with the
hyperk\"ahler cones from the last section. These hyperk\"ahler cones have
$n$ abelian triholomorphic isometries. Any
$4n$ (real) dimensional hyperk\"ahler cone with $n$ such isometries can be
characterized in terms of a homogeneous function $F$, because we can always
dualize back to a description in terms of $n$ tensor multiplets. However,
there may exist equivalent but different tensor multiplet Lagrangians when
there are more than $n$ triholomorphic isometries and one can find
inequivalent subsets of $n$ isometries that commute; we return to this issue
in the next section. According to section 4, the $n$ abelian triholomorphic
isometries of the $4n$-dimensional hyperk\"ahler cone descend to $n$ abelian
quaternionic isometries. This approach gives a complete classification of
all quaternion-\Ka\ spaces of dimension $4(n-1)$ with $n$ quaternionic
abelian isometries as follows: such a space can be dualized by introducing
$n$ tensors, so that one is left with a configuration of $n$ tensors and
$3n-4$ scalars. In the supergravity context, one can argue that this
coupling can be described by a superconformal theory of $n$ tensor
multiplets, which must thus be in the class discussed in the previous
section. Therefore the classification of the quaternionic spaces with $n$
abelian isometries must be complete\footnote{The quotient construction for
the tensor multiplets is subtle for $n=1$ because there are not enough
scalars that can be gauge-fixed; this case has been worked out in
\cite{philippe}.}.

We recall (c.f. \eqn{hyper-KP}) that the hyperk\"ahler potential
$\chi(w,{\bar w},v,{\bar v})=2v^IF_{v^I}(x,v,\bar v)$, where the
coordinates $x^I$ are solved in terms of $w_I+{\bar w}_I,v^I$ and
${\bar v}^I$ by \eqn{def-w}.
The first step in the construction of the quaternion-\Ka\ space is to
reduce to the twistor space ${\cal Z}$ by singling out the $z$ coordinate as
explained in section 3 (cf. \eqn{def-z}). The homothety was given in
\eqn{chi-upper}, and hence we choose
\be
\label{tensor-z}
2v^I\partial_{v^I}\equiv\partial_z\ .
\ee
Since the coordinates on the twistor space have weight zero under
dilatations, it is convenient to use special coordinates (compatible with
\eqn{tensor-z}):
\be
v^{{I}}={\rm e}^{2z}t^{{I}} \,,
\quad \mbox{ with }\,{I}=1,\ldots,n-1\ , \qquad v\equiv v^n={\rm e}^{2z} \ .
\label{spec-coord}
\ee
The twistor space $\cal Z$ is then parametrized  by $2n-1$ complex
coordinates, $t^{I}$, $w_I$, and $w\equiv w_n$, with $I=1,\ldots,n-1$ (here
and henceforth). All these coordinates have zero weight under dilatations. It
is now convenient to change the coordinates $x^I$ of the tensor multiplets
in a similar way,
\be
x^I={\rm e}^{z+\bar z}q^I\ ,
\qquad x\equiv x^n ={\rm e}^{z+\bar z}q\ ,
\ee
where the $q^I$ are real and have zero scaling weight.

Because $F(x,v,{\bar v})$ is homogeneous and invariant under
SO(2), it can be written in terms of a new function ${\cal F}$ as
\be
F(x,v,{\bar v})={\rm e}^{z+\bar z}{\cal F}(q,t,\bar t)\ ,
\ee
which is still restricted by the first and the last
equation of \eqn{inv-F}. We give these restrictions below. The
derivatives with respect to the new
coordinates can be expressed straightforwardly in terms of the old
ones,
\bea
\pa_{x^I}={\rm e}^{-(z+{\bar z})}\pa_{q^I}\ ,
&\quad & \pa_{v^{I}}={\rm e}^{-2z}\pa_{t^{ I}}\ ,
\nonumber\\[2mm]
\pa_{x}={\rm e}^{-(z+{\bar z})}\pa_{q}\ ,  &\quad&
\pa_{v}=\ft12{\rm e}^{-2z}(\pa_z-2t^{ I}\pa_{t^{ I}}
-q^I\pa_{q^I}-q\,\pa_q)\ .
\eea
It follows that the \Ka\ potential of the twistor space is determined
by the function ${\cal F}$,
\be
\label{K-F}
K(t,{\bar t}, w+\bar w)=\ln\Big[{\cal F}(q,t,\bar t)- q^I(w_I+{\bar w}_I)
-q(w+{\bar w})\Big] \ ,
\ee
where the coordinates $q^I$ and $q$ are determined as functions of
$t^I,w_I$ and $w$ from
\be
w_I+{\bar w}_I=\frac{\partial {\cal F}}{\partial q^I}
\,, \qquad w+{\bar w}=\frac{\partial {\cal F}}{\partial q}\ .
\ee
We can also write down the constraints corresponding to the first
equation of \eqn{inv-F}.
\bea
&&{\cal F}_{q^Iq^J}+{\cal F}_{t^I{\bar t}^J} \ =\  0 \ , \nonumber\\
[3mm] \label{lapl-In}
&&{\cal F}_{{ t}^I} + 2{\cal F}_{q^Iq}- 2{\cal F}_{{ t}^I{\bar
t}^J}{\bar t}^J - {\cal F}_{{ t}^Iq^J}q^J-{\cal F}_{t^Iq}\,q\ =\ 0\
,\nonumber \\[3mm]
&&\ft14 {\cal F} -\ft14 (q^I{\cal F}_{q^I}-q{\cal F}_q) + (1+\ft14 q^2)
\,{\cal F}_{qq}  \;    \nonumber \\
&&\quad  +(t^I{\bar t}^J+\ft14 q^I q^J)\,{\cal F}_{q^Iq^J}
+(t^I+{\bar t}^I+\ft12 q^Iq)\, {\cal F}_{q^Iq}\ =\ 0\ .
\eea
The last equation of \eqn{inv-F} implies
\bea
{\cal F}_{q^It^J}-{\cal F}_{q^Jt^I}&=&0\ , \nonumber\\ [3mm]
{\cal F}_{t^Iq}+t^J{\cal F}_{t^Jq^I}+\ft12q^J{\cal F}_{q^Jq^I}+\ft12
q{\cal F}_{q^Iq}&=&0\ .
\eea

In the previous section we discussed how the function $F$ can be
represented by a contour integral (cf. \eqn{contour-F}).
There is a corresponding representation for the function $\cal F$ on
the twistor space ${\cal Z}$. Absorbing a phase factor in the contour
integration variable $\xi$, one can straightforwardly verify the
following expression,
\be
{\cal F}(q,t,\bar t)=\oint\,\frac{{\rm d}\xi}{2\pi i
\xi}\;G(\eta(\xi))\ ,
\ee
where $G$ is still a homogeneous function in the $\eta$'s (up to terms
that vanish under the contour integral). The latter
are now defined by
\be
{\eta}^I(\xi)= \bar t^{I}\,\xi^{-1}  + q^I- {t}^I\, {\xi}\,,\qquad
\eta^n(\xi) = \xi^{-1} + q - \xi \ .
\ee

The holomorphic hyperk\"ahler cone  two-form $\Omega$ was given in
\eqn{tensor-twoform}; in the new coordinates, it becomes
\be
\O={\rm e}^{2z}\Big[2 ({\rm d}w+t^{ I}{\rm d}w_{ I})
\wedge {\rm d}z + {\rm d}w_{ I}\wedge {\rm d}t^{ I}\Big]\ .
\ee
{From} this result one can easily read off the holomorphic one-form $X$
and the holomorphic two-form $\omega$ on the twistor space,
\begin{eqnarray}
X&=&2({\rm d}w+t^{ I}{\rm d}w_{I})\ ,  \nonumber\\
\omega &=&{\rm d}w_{ I} \wedge {\rm d}t^{I} \ .
\end{eqnarray}
By computing the inverse two-form, we find $Y^i$, and hence (see
\eqn{defzeta})
\be
\zeta =  2w\ .
\ee
Decomposing the holomorphic HKC two-form and its
inverse in terms of
the coordinates $(t^I, w_I,\zeta,z)$ one obtains the following results
for the quaternion-\Ka\ manifold $\Quat^{4(n-1)}$:
\be
\omega_{\a\b}= \pmatrix{0&-{\bf 1}\cr {\bf 1}& 0\cr} \,,\qquad X_\a(t)
= (0,2\,t^I)\ , \qquad Z^\a(t) = (2\,t^I, 0)\ .
\ee
Observe that these quantities only depend on $t^I$ and not on $w_I$;
this can be understood from the presence of the triholomorphic
isometries, as discussed below.

At this point one can evaluate the twistor-space metric by taking appropriate
derivatives of the \Ka\ potential \eqn{K-F}. This result can
be expressed in terms of the derivatives of the function $\cal F$, along
the same lines as in the beginning of section 5. For instance, we note
that
\be
K_{t^I} = {\cal F}_{t^I}\, {\rm e}^{-K} \,,\quad
K_{w_I} = -q^I\, {\rm e}^{-K} \,,\quad
K_{\zeta} = -\ft12 q\, {\rm e}^{-K} \ .
\ee

With these result we can write down the quaternion-\Ka\ metric by
following the procedure outlined in section 4. We restrict the indices
to the coordinates $(t^I, w_I)$ and subsequently impose the gauge
condition $\zeta=0$, which implies ${\pa {\cal F}}/{\pa q}=0$.
This equation determines $q$ in terms of the other coordinates, $q=
q(q^I,t^I,{\bar t}^I)$, and hence the scalars of the $n$-th tensor multiplet
are completely eliminated. The different components of the
quaternion-\Ka\ metric are given by
\bea
\label{toric-Q-metric}
G_{w_{ I}{\bar w}_{J}}&=&K_{w_{I}{\bar w}_{J}}
-4{\rm e}^{-2K} t^I{\bar t}^J\ , \nonumber\\
G_{w_{ I}{\bar t}^{ { J}}}&=&K_{w_{ I}{\bar t}^J}\ ,
\nonumber\\
G_{t^{I}{\bar t}^J}&=&K_{t^{ I}{\bar t}^J}\ .
\eea
Likewise we evaluate the quaternion-\Ka\ two-form ${\cal
Q}^+$ defined by \eqn{almost-cs}:
\be
\label{toric-almost-cs}
{\cal Q}^+={\rm e}^{-K}\Big[ {\rm d}w_I\wedge \Big({\rm d}t^I-2t^IK_{t^J}\,
{\rm d}t^J\Big)-2t^IK_{w_J}\, {\rm d}w_I\wedge {\rm d}w_J \Big]\ .
\ee
All the above formulae are evaluated at $\zeta=0$.

We now discuss the abelian isometries.
In terms of the coordinates introduced above, the $n$ commuting
triholomorphic isometries in the hyperk\"ahler cone take the form
\be
\delta w_I=i\alpha_I\ , \qquad \delta w=i\alpha \ ,\qquad
\delta t^I=0 \ ,\qquad \delta z=0\ ,
\ee
where $\alpha$ and $\alpha_I$ are real, constant parameters.
It then follows trivially that the twistor space has the same
set of isometries and that these are holomorphic with respect to the
complex structure on the twistor space. This is in accord with the
general discussion at the end of section 3. We find that the moment map
corresponding to the shift in $w_I$ is simply given by
$\mu_{(I)}=-K_{{\bar w}_I}= -K_{w_I}$, and the one corresponding to the
shift in $w$ is given by $\mu=-K_{\bar w}= -K_{w}$.

In the quaternion-\Ka\ manifold, the shifts in $w_I$ still preserve
the metric \eqn{toric-Q-metric} and thus remain isometries:
$\zeta\equiv 2w$ is not affected by them, and hence the gauge choice
$\zeta=0$ induces no compensating Sp(1) transformation.
This is different for the shift in
$w$, which has to be compensated by an Sp(1) transformation according
to \eqn{Q-isom}. Because the shift in the twistor space acts
exclusively on $\zeta$ and not on any of the other coordinates, the
corresponding isometry in the quaternion-\Ka\ space is directly
proportional to $X^\a$. Hence the  Killing vectors associated with $n$
the quaternionic isometries are given by
\be
k^{w_J}_{(I)}=i\,\delta_J^I\ , \qquad k^\a=i\frac{X^\a}{X^\z}\ ,
\ee
and their complex conjugates. In these formulae one again sets
$\zeta=0$. It follows easily that these isometries
commute, since $X^i$ depends on $w_I+\bar w_I$. Of course,
this is all in accord with the general structure discussed in
section 4.
Likewise the quaternionic structure is invariant
under the shifts in $w_I$, as can be explicitly verified from
\eqn{toric-almost-cs}, while under the $n$-th isometry it rotates
according to \eqn{rot-twoforms}.

\section{The Universal Hypermultiplet}
\setcounter{equation}{0}
In this section we discuss the universal hypermultiplet, which parametrizes a
four-dimensional quaternion-\Ka\ manifold that appears as part of the moduli
space of Calabi-Yau compactifications of type II strings. Classically the
relevant homogeneous space is \cite{cfg}
\be
\Quat_{{\rm UH}}=\frac{\rm U(1,2)}{{\rm U}(1)\times {\rm U}(2)}\
,\label{Q-UHM}
\ee
which is the lowest dimensional case of the Wolf spaces
${X(n-1)}$. These spaces, which are homogeneous quaternion-K\"ahler manifolds
of real dimension ${4(n-1)}$, are given by \cite{Wolf},
\be
X(n-1)=\frac{{\rm U}(n-1,2)}{{\rm U}(n-1)\times {\rm U(2)}} \ .
\ee
The $X(n-1)$ belong to the class of
\Ka\ spaces $G(p,q)$ of real dimension $2pq$,
\be
G(p,q) ={ {\rm U}(p,q)\over {\rm U}(p) \times {\rm U}(q)}\,,
\ee
called Grassmannian manifolds.
To exhibit the distinction between the parametrization
of $X(n-1)$ as a quaternion-\Ka\ and as a \Ka\ manifold, let us briefly
discuss the coset representative for the $G(p,q)$ and their
corrsponding \Ka\ potentials.

We start with the coset representative (see {\it e.g.}, \cite{Gilmore}),
decomposed as follows,
\be
{{\rm U}(p,q)\over {\rm U}(p)\times {\rm U}(q)} = \pmatrix{ \sqrt{{\bf
1}_p \mp X X^\dagger} & X\cr \noalign{\vskip2mm}
\mp X^\dagger & \sqrt{{\bf 1}_q \mp X^\dagger X} \cr}\,,
\label{coset-repr}
\ee
where $X$ is a $p$-by-$q$ complex matrix and the upper(lower) sign describes
the compact(noncompact) version; in the compact case, ${\rm U}(p,q)$ is
replaced by ${\rm U}(p+q)$. The $pq$ complex parameters
contained in $X$ provide (local) coordinates on $G_{(p,q)}$.

To exhibit the \Ka\ structure, it is convenient to adopt inhomogenous
coordinates defined by $Z = X\,[ {{\bf 1}_q \mp X^\dagger
X}]^{-1/2}$. Then the line element takes the form,
\be
{\rm d}s^2 = \pm \,{\rm Tr}\left[ {1\over { {\bf 1}_p \pm Z
Z^\dagger}} \,{\rm d}
Z\, {1\over {{\bf 1}_q \pm Z^\dagger Z}} \, {\rm d} Z^\dagger \right] \;.
\ee
Clearly the metric is hermitean in the coordinates $Z$ and
$Z^\dagger$; it has a \Ka\ potential,
\be
K(Z,Z^\dagger) = {\rm Tr} \Big[ \ln [{\bf 1}_p \pm Z Z^\dagger]
\Big]\;.  \label{G-K-potential}
\ee
However, as we shall see in the next subsection, the coordinates $Z$
are not the appropriate coordinates for the quaternion-\Ka\
description of $X(n-1)$.

Returning to the universal hypermultiplet space $\Quat_{\rm UH}$ we
mention that one finds different expressions for its K\"ahler
potential in the literature. One is based on complex
coordinates $S$ and $C$, which
refer to the dilaton/axion complex and the R-R fields, with
\be
\label{K-S}
K_{{\rm UH}}=\ln(S+{\bar S}-2C{\bar C})\ .
\ee
Another one, based on coordinates $u$ and $v$ and K\"ahler potential
\be
\label{K-u}
K_{{\rm UH}}=\ln(1-u{\bar u}-v{\bar v})\ ,
\ee
corresponds directly to the parametrization  \eqn{G-K-potential}.
The two potentials are related by a holomorphic coordinate transformation
\be
S=\frac{1-u}{1+u}\ , \qquad
C=\frac{v}{1+u}\ ,
\ee
and a \Ka\ transformation $\ln[(1+u)/\sqrt{2}] + h.c.$.

We now describe $\Quat_{{\rm UH}}$ starting from an
appropriate hyperk\"ahler cone, using the techniques of this paper. In
the first subsection, we construct this hyperk\"ahler cone and its
corresponding twistor space for all $X(n-1)$, and specialize at the end
to $n=2$. In a second subsection, we rederive the same geometries using the
Legendre transform method and the contour integral formalism of section 5.
The hyperk\"ahler geometry associated to the Wolf spaces has also been
discussed in the mathematics literature \cite{KS}.

\subsection{Quotient construction of $X(n-1)$}
It is well known \cite{BreitSohn,Galicki2} that the Wolf spaces
$X(n-1)$ can be obtained by performing a hyperk\"ahler quotient
of a flat space of complex dimension $2n+2$ followed by an $N=2$
superconformal quotient (the two quotients can be performed in opposite
order: first the $N=2$ superconformal quotient followed by a quaternionic
one \cite{Galicki2}). In this section we first perform the hyperk\"ahler
quotient and obtain a $4n$-dimensional hyperk\"ahler cone and the
corresponding twistor space ${\cal Z}$. Subsequently we obtain the
quaternion-\Ka\ space via the procedure outlined in section 4. Thus we start
by considering ${\bf C}^{2n+2}$, with a pseudo-Riemannian metric with $n+1$
pairs of complex coordinates denoted by $(z_+^I,z_{-I})$. We can distinguish
a number of obvious symmetry groups that act linearly on these coordinates.
First, $z_+$ and $z_-$ transform in the (inequivalent) conjugate fundamental
representations of U$(n-1,2)$ (or its compact version) so that their product
is invariant. Hence, when $z_+$ transforms as $z_+^I\to U^I{}_{\!J}
\,z^J_+$, then $z_-$ transforms according to $z_{-I}\to (U^{-1})^J{}_{\!I}
\,z_{-J}$. The noncompact versions of U$(n+1)$ satisfy $(U^{-1})^I{}_{\!J} =
\eta^{I\bar I} \,\bar U^{\bar J}{}_{\bar I}  \,\eta_{J\bar J}$, where
$\eta_{I\bar J}$ is a diagonal matrix with entries $\pm1$, and $\eta^{I\bar
J}$ is its inverse. Furthermore, $z_+$ and $\eta \bar z_-$ transform as a
doublet under SU(2); this is the Sp(1) that rotates the complex structures
\eqn{rot-J}. The above assignments under ${\rm U}(n-1,2)\times {\rm SU}(2)$
are characteristic for the coset-space structure of $X(n-1)$, but the
holomorphic assignements are different. To be specific, the matrix $X$ in
the coset representative (cf. \eqn{coset-repr} with $p=n-1$ and $q=2$) is
related to $(z_+^I,\eta^{I \bar J} \bar z_{-J})$.

The flat space ${\bf C}^{2n+2}$ is obviously a  hyperk\"ahler cone,
and its hyperk\"ahler potential is
\be
\chi_{{\rm (2n+2)}}=\eta_{I\bar J} \,z^I_+{\bar z}^J_+ + \eta^{I\bar
J}\, z_{-I} {\bar z}_{-J} \ .\label{chi-flat}
\ee
We assume that the last two eigenvalues of $\eta_{I\bar J}$ are positive.
For the noncompact spaces the remaining entries are
equal to $-1$, while for the compact spaces $\eta$ is the unit matrix.
Furthermore, the U$(n-1,2)$ invariant holomorphic two-form is given by
\be
\Omega ={\rm d}z_+^I \wedge {\rm d}z_{-I} \ . \label{flat-two}
\ee

The isometry we quotient ${\bf C}^{2n+2}$ by is the U(1)
subgroup of ${\rm U}(n-1,2)$ which
acts on $z_+$ and $z_-$ with opposite phase. The three moment maps
associated with this isometry are
\be
\mu^+ =-i z_+^I\, z_{-I} \,,\qquad \mu^- = \bar \mu^+ \,,\qquad \mu^3 =
- \eta_{I\bar J}\, z^I_+{\bar z}^J_+
+ \eta^{I\bar J}\,z_{-I}{\bar z}_{-J} \ . \label{moment-U1}
\ee
The hyperk\"ahler quotient \cite{F2,Hitchin} is performed by gauging
the isometry,
integrating out the corresponding connection and setting the moment
maps to zero. Note that this hyperk\"ahler quotient is
consistent with ${\rm U}(n-1,2)\times {\rm SU}(2)$.

One can now proceed in two equivalent ways. One is to introduce
gauge-invariant inhomogeneous coordinates $z^I_+/z_+^{n+1}$ and
$z_{-I}z_+^{n+1}$. Because of U(1) gauge invariance the phase of $z_+^{n+1}$
drops out while the vanishing of the moment maps implies that $\vert
z_+^{n+1}\vert$ and $z_{-{n+1}}z_+^{n+1}$ are constrained in terms of the
remaining (inhomogeneous) coordinates. We are thus
left with only $n$ coordinates $z_+^a$ and $z_{-a}$ with $a =
1,\ldots, n$. The resulting space is still a hyperk\"ahler cone whose
potential (now in terms of the inhomogeneous coordinates) equals
\be
\chi_{(2n)} = 2\,\chi_+\,\chi_-\ ,
\ee
where
\be
\chi_+ = \sqrt{\eta_{I\bar J}\, z^I_+{\bar z}^J_+ }\;, \qquad
\chi_-= \sqrt{\eta^{I\bar J}\, z_{-I}{\bar z}_{-J}} \ ,
\ee
and
\be
z_+^{n+1} = 1\,,\qquad z_{-{n+1}} = -  z_+^a \,z_{-a} \ .
\ee
The dilatations act on $z_+$ and $z_-$ with scaling weights
$0$ and $2$, respectively. Furthermore the holomorphic two-form
\eqn{flat-two} takes the form
\be
\Omega= {\rm d}z^a_+\wedge {\rm d}z_{-a}
\ee
on the quotient.

The same results follow using the $N=1$ superspace formalism to gauge the
U(1) isometry. The hyperk\"ahler potential \eqn{chi-flat} is gauged to
\be
{\hat \chi}_{{\rm (2n+2)}}={\rm e}^V\eta_{I\bar J} \,z^I_+{\bar z}^J_+ +
{\rm e}^{-V}\eta^{I\bar
J}\, z_{-I} {\bar z}_{-J} \ .\label{chi-gauged}
\ee
The (anti)holomorphic moment maps $\mu^{\pm}$ are unchanged, and $\mu^3$
becomes
\be
{\hat \mu}^3 =
- {\rm e}^V\eta_{I\bar J}\, z^I_+{\bar z}^J_+
+ {\rm e}^{-V}\eta^{I\bar J}\,z_{-I}{\bar z}_{-J} \ .
\ee
Now we may solve ${\hat \mu}^3=0$ for $V$ and substitute back into
\eqn{chi-gauged}; the geometric meaning of this procedure
is explained in \cite{Hitchin}. In $N=1$ superspace, the
gauge group is complexified and hence we may choose the gauge
$z_+^{n+1}=1$, while $z_{-{n+1}}$ is again determined by the
holomorphic momentum map constraint. This is all derived
from $N=2$ superspace in appendix B.

Finally we note that the SU(2) transformations that rotate the
complex structures \eqn{rot-J} of the quotient take the form
\bea
\d z_+^a &=& \epsilon^+\,\frac{\chi_+}{\chi_-}\,
\Big[ \eta^{a\bar b} +z_+^a \,\bar z_+^b\Big] \bar z_{-b} \ ,\nonumber \\
\d z_{-a} &=& - \epsilon^+ \Big[ \frac{\chi_-}{ \chi_+}\,
\eta_{a\bar b}+\frac{\chi_+}{\chi_-}\,z_{-a} \,\bar z_{-b} \Big]
\bar z_+^b \ .
\eea
The SU(2) invariance of the
hyperk\"ahler potential can explicitly be checked using
\be
\d_{\epsilon^+} \chi_\pm = \pm  {\epsilon^+ \, \chi^2_+\,\bar z_+^a\bar
z_{-a}\over 2 \chi_\mp}\ .
\ee
A similar analysis can be done for the group of  SU$(n-1,2)$ triholomorphic
isometries. For finite transformations these take the form,
\bea
z_+^a &\to& {U^a{}_{n+1} + U^a{}_b \, z_+^b \over  U^{n+1}{}_{n+1}
+ U^{n+1}{}_c \, z_+^c} \ , \nonumber \\[2mm]
z_{-I} &\to& [(U^{-1})^J{}_I \, z_{-J}] \,
[U^{n+1}{}_{n+1} + U^{n+1}{}_a \, z_+^a]  \ . \label{su(n-1,2)}
\eea
The factors $\chi_\pm$ of the hyperk\"ahler potential transform as
\be
\chi_\pm \to \vert U^{n+1}{}_{n+1} +
U^{n+1}{}_a \,z_+^a\vert^{\mp 1}\, \chi_\pm\ ,
\ee
so that $\chi$ is indeed invariant.

Now we descend to the twistor space, decomposing the coordinates as
\be
z_{-n} = {\rm e}^{2z}\ ,\qquad z_{-i} =  {\rm e}^{2z}\,u_i \qquad(i=
1,\ldots,n-1) \ .
\ee
In terms of these coordinates the holomorphic two-form is
\be
\Omega=2\,{\rm e}^{2z}\,({\rm d}z_+^n+u_i\,{\rm d}z^{i}_+)\wedge
{\rm d}z+  {\rm e}^{2z} \,
{\rm d}z_+^{i}\wedge {\rm d}u_i \ .\label{hol-2form-uhm}
\ee
{From} this we can read off the holomorpic one-form on the twistor
space,
\be
X=2\, {\rm d}z_+^n+ 2\,u_i\,{\rm d}z_+^{i}\ ,\label{X-UHM}
\ee
which shows that the coordinate $\zeta$ is given by $2z_+^n$. Hence we
identify
\be
2z_+^n = \zeta\ ,\qquad z_+^{i} = v^i\ ,
\ee
so that the holomorphic two-form is
\be
\Omega = {\rm e}^{2z} \Big[( {\rm d}\zeta + 2 u_i\,{\rm d}v^i )\wedge
{\rm d}z +  {\rm d}v^{i}\wedge {\rm d}u_i\Big] \ .
\ee
The \Ka\ potential of the twistor space is given by
\be
K( u,v, \zeta,  \bar u,\bar v, \bar\zeta) = \ln \left[ 2
\,\chi_+(v,\zeta,\bar v,\bar \zeta)  \;\chi_-(u,v,\zeta,\bar u, \bar
v,\bar \zeta)\right] \ ,
\ee
with
\bea
\chi_+(v,\zeta, \bar v,\bar \zeta)
&=& \sqrt{1 +\ft14  \zeta\bar \zeta + \eta_{i\bar\jmath}\, v^i\bar
v^j}\;, \nonumber \\[2mm]
\chi_-(u,v,\zeta,\bar u,  \bar
v,\bar \zeta) &=& \sqrt{1+ \vert \ft12 \zeta+ u_iv^i\vert^2 +
\eta^{i\bar\jmath}\, u_i\bar u_j} \ .
\eea
The action of the triholomorphic ${\rm SU}(n-1,2)$ isometries follows
straightforwardly from \eqn{su(n-1,2)} and the definitions of the new
coordinates $u,v,\zeta$. All of this is in accord with the analysis
presented in section~3.

We now have all the ingredients to construct the metric of the
quaternionic spaces $X(n-1)$. To avoid lengthy formulae, we return
to the case $n=2$, where we can drop the indices $i,j$. We
simultaneously treat both the compact and the noncompact case by
allowing $\eta=\pm 1$. Imposing the gauge condition $\zeta=0$,
the twistor space \Ka\ potential becomes
\be
K( u,v,0 , \bar u,\bar v, 0)= \ft12 \ln\Big[ 1\pm  u\bar u  ( 1
\pm v \bar v) \Big] +\ft12 \ln \Big[1 \pm v\bar v\Big] + \ln 2\ .
\ee
Using \eqn{X-UHM},
the metric then follows from \eqn{QK-metric}:
\bea
G_{u{\bar u}}&=&\pm\, \frac{1\pm v{\bar v}}{2\,[1\pm u\bar u(1 \pm v\bar
v)]^2}\ , \nonumber\\[2mm]
G_{u\bar v}&=&\frac{\bar u v}
{2\,[1\pm u\bar u(1 \pm v\bar v)]^2} \ ,  \nonumber\\[2mm]
G_{v\bar v}&=&\pm\,\frac{1\pm u\bar u(1\pm v\bar v)^2}
{2\,[1\pm u\bar u(1\pm v\bar v)]^2[1\pm v\bar v]^2}\ .
\eea
Similarly, the quaternion-\Ka\ two-form ${\cal Q}^+$ follows from
\eqn{QK-forms},
\be
{\cal Q}^+= { {\rm d}v\wedge {\rm d}u\over 2\,[1\pm u\bar u(1\pm v\bar
v)]^{3/2}[1\pm v\bar v]^{1/2} } \ .
\ee

Although the metric is obviously hermitean, it is not manifestly
K\"ahler. We already noted that the holomorphic assignments are
different for the \Ka\ and quaternion-\Ka\ formulations, so there must be
a {\it nonholomorphic} coordinate transformation to a coordinate system
in which the metric is manifestly K\"ahler:
$G_{\alpha \bar \beta}=\partial_\alpha \partial_{\bar \beta} K_{{\rm UH}}$
for some \Ka\ potential $K_{{\rm UH}}$.
It is not difficult to find this coordinate transformation,
\be
u^\prime=u(1\pm v\bar v)\,, \qquad v^\prime=\bar v\ ,
\ee
and the corresponding \Ka\ potential takes the standard form,
\be
K_{{\rm UH}}=\ft12\,{\rm ln}(1\pm u^\prime\bar u^\prime \pm v^\prime\bar
v^\prime)\ ,
\ee
in accordance with \eqn{G-K-potential}.
Note, however, that the quaternion-\Ka\ two-form $\Q^+$ is no longer
a $(2,0)$-form in these coordinates.

\subsection{Tensor multiplet description and dualities}

In this subsection, we discuss how the universal hypermultiplet
can be obtained from tensor multiplets via the Legendre transform method
of sections 5 and 6.  We seek a function $F$ that
satisfies the constraints \eqn{inv-F}, and yields the correct
hyperk\"ahler potential for the hyperk\"ahler cone related to the universal
hypermultiplet. We first treat the whole class of
spaces $X(n-1)$. As explained at the end of section 5, the
function $F$ can be represented as a contour integral,
\be
F(x^I, v^I,\bar v^I) ={\rm Re}\oint\,\frac{{\rm d}\xi}{2\pi i
\xi}\,G(\eta(\xi))\ ,
\ee
where
\be
\eta^I(\xi)={\bar v}^I \,\xi^{-1} + x^I - v^I\,\xi \ ,
\ee
and $G$ is homogeneous of first degree under the contour integral.

For a free hypermultiplet, the function $G$ is given by
\eqn{H-eta} and integration over a suitable contour yields the
function $F$ given in \eqn{F-HPn}. Introducing an
arbitrary sign factor $\s$ in front of the functions $F$ and $G$, such as
to allow for a pseudo-Riemannian metric, the identification
of the coordinates is given by
\be
z_{\pm}={\rm e}^{\pm w}\sqrt v\,, \qquad  x=- 2 \s\, {\sqrt {v\bar
v}}\,\sinh (w+\bar w)\ . \label{substitution-vx}
\ee
We now consider $n+1$ free hypermultiplets, so that the function
$F$ is
\be
F_{\rm (2n+2)}=\sum_{I=1}^{n+1} \s_I F^I\ ,
\ee
where $F^I$ is the function \eqn{F-HPn} for the $I'th$ tensor multiplet:
\be
F^I\equiv F_2(x^I,v^I,{\bar v}^I)=r^I-x^I[ \ln(x^I +r^I)-
\ft12 \ln(4\,v^I{\bar v}^I)]\ ,
\ee
where $r^I={\sqrt {(x^I)^2+4v^I{\bar v}^I}}$. This describes
${\bf C}^{2n+2}$; the $\s_{I}= \pm 1$
are the signature factors introduced
before (in the previous subsection, they are $\eta_{I\bar I}$).
The signs associated with $I=n,n+1$ are again taken
positive. Note that the $F^I$ are even functions under $x\to -x, v\to -v$,
modulo terms which do not contribute to the action. Indeed, the second
derivatives of $F_2$ are symmetric with respect to
this uniform sign change (see \eqn{F''}) and the general discussion in
appendix A). The moment maps \eqn{moment-U1} follow now by direct
substitution of
\eqn{substitution-vx} and are given by \cite{Hitchin}
\be
\mu^+ = -i\,\sum_I v^I \,,\qquad \mu^3 =  \sum_I x^I \ .
\ee
There is no U(1) gauge symmetry acting on the tensor multiplet, so
the hyperk\"ahler quotient amounts to imposing the three constraints
$\mu^\pm=\mu^3=0$. This leads to the elimination of  $x^{n+1}$ and
$v^{n+1}$, and we obtain
\be
F_{{\rm (2n)}}=\sum_{a=1}^n\s_a F^a
+F_2(\sum_a x^a,\sum_a v^a,\sum_a {\bar v}^a)\ .
\ee
For $n=2$, we thus find
\be
F_{{\rm UH}} = \s F_2(x^1,
v^1,\bar v^1) + F_2(x^2, v^2,\bar  v^2) + F_2(x^1+x^2, v^1+v^2,\bar v^1+
\bar v^2) \ .
\ee
After performing the Legendre transform, we find the eight-dimensional
hyperk\"ahler cone whose corresponding quaternion-\Ka\ space is equal
to $X(1)$ or $CP^2$ for $\s=-1$ and $\s=1$, respectively.

The function $F_{{\rm UH}}$ can be represented as a contour
integral of\footnote{The signs in this equation are somewhat symbolic, as
they only make sense after the orientation of the contour is determined, and
as each of the three terms are integrated along different contours. Note
that whereas $F_{{\rm UH}}$ is an even function as discussed above, $G_{{\rm
UH}}$ appears to be odd; this apparent discrepancy is absorbed by
a change in the orientation of the contour.}
\be
G_{{\rm UH}}(\eta^1,\eta^2)= \s \eta^1 \ln\eta^1 + \eta^2 \ln \eta^2 +
(\eta^1+\eta^2)\ln (\eta^1+\eta^2) \ .\label{G1}
\ee

As mentioned in the beginning of section 6,
there are actually different contour integral representations for the
same hyperk\"ahler space. This can be understood as follows: On any
hyperk\"ahler manifold, a hypermultiplet can be dualized to a tensor
multiplet whenever the manifold has a triholomorphic isometry.
When the manifold has $n$ {\it commuting} triholomorphic isometries,
one can dualize $n$ hypermultiplets to tensor multiplets. It may happen
that the manifold has a non-abelian triholomorphic isometry group
which contains inequivalent sets of $n$ commuting isometries.
Dualizing with respect to the different sets gives rise to different
tensor multiplet actions. This situation arises for the
universal hypermultiplet. As discussed in the previous subsection,
the hyperk\"ahler cone above $\Quat_{{\rm UH}}$ has an SU(1,2) group of
triholomorphic isometries. Because this group is noncompact, there
are (at least) three inequivalent sets of commuting pairs of
generators. The first set has two compact abelian isometries; we can choose
the generators
\be
T_1=\pmatrix {-1 & 0 & 0 \cr 0 & -1 & 0 \cr 0 & 0 & 2}\ ,
\qquad
T_2=\pmatrix {-1 & 0 & 0 \cr 0 & 1 & 0 \cr 0 & 0 & 0}\ .
\label{gen1}
\ee
The second set has one compact and one noncompact (nilpotent) isometry with
generators
\be
T_1=\pmatrix {-1 & 0 & 0 \cr 0 & -1 & 0 \cr 0 & 0 & 2}\ ,
\qquad
T_2^\prime =\pmatrix{1 & 1 & 0 \cr -1 & -1 & 0 \cr 0 &0 & 0}\ ,
\label{gen2}
\ee
and the third set has two noncompact isometries, nilpotent of order three
and two, respectively, with generators
\be
T_1^\prime=\pmatrix {0 & 0 & -1 \cr 0 & 0 & 1 \cr 1 & 1 & 0}\ ,
\qquad
T_2^{\prime}=\pmatrix {1 & 1 & 0 \cr -1 & -1 & 0 \cr 0 & 0 & 0}\ .
\label{gen3}
\ee
As explained in appendix B, the first set gives rise to \eqn{G1},
the second set gives
\be
G_{{\rm UH}}=\eta^1\,{\rm ln}\,\frac{\eta^1}{\eta^2}\ ,\label{G2}
\ee
whereas the third set gives
\be
G_{{\rm UH}}=\frac{(\eta^1)^2}{\eta^2}\ .\label{G3}
\ee
Clearly these functions are homogeneous of first
degree, and hence correspond to hyperk\"ahler cones. The last form also
appeared in \cite{Siegel}.

The functions $F(x,v,\bar v)$ associated with these expressions for
$G(\eta)$ take a rather complicated form; the functions
$G(\eta)$ are the most concise way to encode the structure of the tensor
Lagrangians and the corresponding hyperk\"ahler cones. This suggests that we
should try to understand the physics in directly terms of $G(\eta)$, rather
than in terms of the corresponding HKC or quaternion-\Ka\ space.

\section{Discussion and conclusion}
\setcounter{equation}{0}
In this paper, we have explored the relation between hyperk\"ahler
cones and quaternion-\Ka\ geometries. This relationship is one-to-one:
every hyperk\"ahler cone has an associated quaternion-\Ka\ manifold and
vice versa. From the superconformal multiplet calculus in supergravity
\cite{DVV} it was known how to associate a quaternion-\Ka\ space with a
hyperk\"ahler cone \cite{DWKV} by the $N=2$ superconformal quotient. Here
we have found explicit convenient coordinates on the quaternion-\Ka\
manifold by making appropriate gauge choices; along the way, we
constructed the explicit \Ka\ potential on the twistor space of the
quaternion-\Ka\ space. Furthermore, we have used the relation to tensor
multiplets to classify $4(n-1)$-dimensional quaternion-\Ka\ geometries with
$n$ commuting quaternionic isometries. In principle one can also find the
hyperk\"ahler cone corresponding to any quaternion-\Ka\ manifold; although
no uniform and explicit description has been given in the supergravity
context, this has been shown in the mathematics literature
\cite{Swann,Galicki}.

Hyperk\"ahler cones and quaternion-\Ka\ spaces
have many applications in physics. Yang-Mills instanton moduli spaces in
four Euclidean dimensions have this geometrical structure. The
conformal symmetry of the four-dimensional spacetime is carried over to the
moduli space of the collective coordinates, so that the size of the
instanton is the cone variable. In particular, the one-instanton moduli
spaces corresponding to the simple gauge groups are cones over the (compact)
Wolf spaces \cite{Vandoren}; the specific spaces $X(n-1)$ discussed in
section~7 appear as moduli spaces of a single ${\rm SU}(n+1)$ instanton. An
explicit representation has been written down for the hyperk\"ahler
potential $\chi$ associated with the instanton moduli spaces
\cite{Maciocia},
\be
\chi = \int {\rm d}^4x \; x^2 \, {\rm Tr}\, F_{\m\n}^2 \ ,
\ee
which is a function of the collective coordinates of the
instanton solution. It would be interesting to determine the
eight-dimensional quaternion-\Ka\ geometry for the centered two-instanton
solution in SU(2).

The moduli space of a Calabi-Yau compactification of type-II string theory
also has a quaternion-\Ka\ sector which contains the dilaton. Here, few
results are known for the full quantum moduli space.
At the perturbative level \cite{quat-quantum},
the space has a large number of continuous abelian isometries associated
with the axion and the R-R fields. Beyond the perturbative level
\cite{nonpert}, one expects that there are no continuous isometries and
only a number of discrete isometries may survive (much as
the $\theta$-angle in quantum-chromodynamics cannot be shifted
by an arbitrary constant because of instanton corrections).

The following analogy \cite{hyp-3d} suggests that our results may lead to a
determination of the quantum corrections in moduli spaces parametrized by
hypermultiplets: The centered moduli space of two SU(2) monopoles \cite{AH},
or equivalently, the Coulomb branch of three-dimensional $N=4$ SU(2)
Yang-Mills theory \cite{sw3d}, is hyperk\"ahler. The asymptotic (or
perturbative) moduli space has a triholomorphic isometry, and is easily
described in terms of an $N=4$ vector multiplet, which is equivalent to a
tensor multiplet (an $O(2)$ multiplet in the terminology of appendix B, see
\eqn{o2mult}) in three dimensions. Nonperturbative corrections break this
isometry. In the contour integral representation discussed in sections 5 and
7 and appendix B, the integrand $G(\eta(\xi))$ (which is the projective
superspace Lagrangian \eqn{eq-action} of appendix B) is
essentially unchanged; rather, the multiplet changes to an $O(4)$
multiplet \eqn{o4mult} \cite{itimr}. A remarkable feature of this mechanism
is that the $O(4)$ description automatically incorporates all
nonperturbative corrections with no adjustable parameters. The
rigidity of the conformal structure of hyperk\"ahler cones leads us to
speculate that a similar miracle may occur for the Calabi-Yau moduli
spaces.

\vspace{8mm}

\noindent
{\bf Acknowledgement}\\
We are happy to acknowledge useful discussions with N. Berkovits, K. Galicki,
C. Lebrun, and W. Siegel.  Some of this work was done with support from NSF
grant No.\ PHY-9722101. B.d.W. thanks the Aspen Center for Physics for its
hospitality during the workshop ``String Dualities and their Applications''.
S.V. thanks the Erwin Schr\"odinger Institute for hospitality during the
workshop ``Holonomy Groups and Differential Geometry''.

\appendix
\section{Superconformal constraints on $F(x,v,\bar v)$}
\label{app-A}
\setcounter{equation}{0}
In the text we argued that conformal invariance of the Lagrangian
\eqn{tensor-L} requires that $F_{x^Ix^J}$,
and $F_{x^Iv^J}$ be homogeneous of degree $-1$ and covariant
under phase transformation of the coordinates $v^I$. Hence we
expect that $N=2$ superconformal invariance
requires \eqn{laplace} and the following four
identities (and their complex conjugates):
\bea
x^K F_{x^I x^J x^K} +
v^K F_{x^I x^J v^K} +
\bar v^K  F_{x^I x^J \bar v^K} &=& - F_{x^I x^J}\ ,\nonumber\\
x^K F_{x^I v^J x^K} +
v^K F_{x^I v^J v^K} +
\bar v^K F_{x^I v^J \bar v^K} &=& - F_{x^I v^J}\ ,\nonumber\\
v^K F_{x^I x^J v^K} -
\bar v^K F_{x^I x^J \bar v^K} &=& 0 \ ,\nonumber\\
v^K F_{x^I v^J v^K} -
\bar v^K F_{x^I v^J \bar v^K} &=& - F_{x^I v^J}\ .
\label{homo-so2}
\eea
These constraints imply that certain derivatives of the function
$F(x,v,\bar v)$ are homogeneous functions of degree $-1$ and covariant under
phase transformations of the coordinates $v^I$ and $\bar v^I$.

We now derive the consequences of \eqn{homo-so2}. We use \eqn{laplace}
throughout this appendix. The first two equations \eqn{homo-so2} imply
that
\be
x^I F_{x^Ix^J} + v^I F_{v^Ix^J} + \bar v^I F_{\bar v^Ix^J} = c_J\ ,
\ee
with $c_J$ some real integration constants. Integrating
once more shows that $F$ can be decomposed according to
\be
F(x,v,\bar v) = F^\prime(x,v,\bar v) + c_Jx^J \,\ln c_Ix^I + f(v,\bar
v) \ ,
\ee
where $F^\prime$ is a homogeneous function of first
degree and $f(v,\bar v)$ is any function independent of $x$.

Along similar lines one establishes that
\be
x^I F_{x^I} + v^I F_{v^I} + \bar v^I F_{\bar v^I} -F  = g(v) + h(\bar
v, x)\ ,
\ee
where $g(v)$ and $h(\bar v,x)$ are again some unknown
functions. From the fact that the right-hand side of this equation
must be real we deduce that $h(\bar v,x)$ equals the sum
of $\bar g(\bar v)$ and some function of $x^I$. Combining all
information we thus find that $F(x,v,\bar v)$ can be written as
follows,
\be
F(x,v,\bar v) = F^{\prime\prime}(x,v,\bar v) + \ft12 c_Jx^J \,\ln\vert
c_Iv^I\vert^2 + G(v) + \bar G(\bar v) \ ,
\ee
where $F^{\prime\prime}(x,v,\bar v)$ is a homogeneous function of
first degree (related to $F^\prime$) and satisfies \eqn{laplace}.
Observe that the Lagrangian \eqn{tensor-L} for the
tensor multiplets only receives contributions from
$F^{\prime\prime}$ and not from the other terms on
the right-hand side, so that those can be dropped\footnote{Equivalently,
these terms are total derivatives in $N=1$ superspace.}.

We now analyze the last two equations of
\eqn{homo-so2}. Because all terms other than $F^{\prime\prime}$
already satisfy these equations, we may restrict our attention to the
homogeneous function and replace $F$ by $F^{\prime\prime}$. We then
prove that
\be
v^I F^{\prime\prime}_{v^Ix^J} - \bar v^I F^{\prime\prime}_{\bar
v^Ix^J} = i \tilde c_J\ ,
\ee
with the $\tilde c_J$ some real integration constants. Furthermore,
using \eqn{laplace} we find
\be
{\pa\over\pa v^J}{\pa\over\pa \bar v^K}\Big(v^I F^{\prime\prime}_{v^I}
- \bar v^I F^{\prime\prime}_{\bar v^I} \Big) = 0\ .
\ee
These two results lead to the following decomposition of the function
$F$,
\be
\label{totalF}
F(x,v,\bar v) = F_{\rm inv}(x,v,\bar v) + \ft12 c_Jx^J \ln\vert
c_Iv^I\vert^2 + \ft12i\tilde  c_Jx^J \ln(\tilde c_Iv^I/ \tilde
c_K\bar v^K) +\ G(v) + \bar G(\bar v) \ ,
\ee
where $F_{\rm inv}(x,v,\bar v)$ is a homogeneous function of first
degree which is invariant under phase transformations of the $v^I$ and
satisfies \eqn{laplace}, and up to total derivatives, determines the action.

We can further restrict $F_{\rm inv}$ such that
$F_{x^Iv^J}$ is symmetric in $I$ and $J$, as claimed in \eqn{inv-F}, by
using the freedom to shift $F_{\rm inv}$ by terms
of the form $x^I(f_I(v)+{\bar f}_I({\bar v}))$,
which do not contribute to the action. Here the $f_I(v)$ are arbitrary
homogeneous and holomorphic functions of degree zero.
Observe that, using \eqn{laplace}, the antisymmetric part
\be
a_{IJ}\equiv F_{x^Iv^J}-F_{x^Jv^I}\ ,
\ee
is independent of $x$ and $\bar v$, and hence is purely holomorphic in
$v$. Moreover, the two-form $a$ is closed, and is therefore locally
exact as a function of $v$, {\it i.e.}, $a_{IJ}=\pa_I a_J(v)-\pa_J a_I(v)$.
It is then clear that if we redefine $F_{\rm inv}$ by choosing $f_I=a_I$,
the new $F_{x^Iv^J}$ is symmetric.

All the remaining terms in the function $F$ in \eqn{totalF}
give rise only to total derivatives in the Lagrangian \eqn{tensor-L}.
Therefore they can be ignored, and we restrict $F(x,v,\bar v)$ to be
homogeneous of first degree, U(1) invariant, with a symmetric $F_{x^Iv^J}$
and subject to \eqn{laplace}. The results lead to \eqn{inv-F}.

\section{Projective superspace and tensor multiplet dualities}
\label{app-B}
\setcounter{equation}{0}
In this appendix, we review the projective superspace 
formalism\footnote{A related formalism has been developed in harmonic
superspace; for some references, see \cite{hss}.} for $N=2$
supersymmetry \cite{proj-ss,KLR,lr} and use it prove the equivalence between
the three different
contour integral representations of the universal hypermultiplet given
by \eqn{G1}, \eqn{G2} and \eqn{G3}.

\subsection{Projective superspace}
This subsection is taken essentially verbatim from \cite{PSS}.

The algebra of $N=2$ supercovariant derivatives in four dimensions
is
\be
            \{ D_{i \a} , D_{j \b} \} = 0 \ ,\qquad
            \{ D_{i \a} , \bar{D}^j_{\db} \} = i \d^j_i \pa_{\a\db } \  ,
\label{n2_algebra}
\ee
where $i,j=1,2$ are SU(2) isospin indices and $\a,\db=1,2$ are Lorentz
spinor indices.  We define an abelian subspace of $N=2$ superspace
parameterized by a complex projective
coordinate $\xi$ and spanned by the supercovariant derivatives
\be
            \na_\a (\xi)  = D_{1 \a} + \xi D_{2 \a}\ , \qquad
            \bar{\na}_{\da} (\xi)  =  \bar{D}^2_{\da} - \xi \bar{D}^1_{\da}\ .
\ee
For notational simplicity we write $D_{1 \a} =D_\a, D_{2 \a} = Q_\a$.
The conjugate of any object is constructed in this subspace by
composing the antipodal map on the Riemann sphere with
hermitean conjugation $\xi^{\ast}\to-1/\xi$ and multiplying by an
appropriate factor. For example,
\be
\bar{\na}_{\da} (\xi) = \(- \xi \) \( \na_\a \)^{\ast} \(-{1 \over \xi}\)
    = \(- \xi \) \( \bar{D}^{1}_{\da} + \(-{1 \over \xi}\)
\bar{D}^{2}_{\da}\)
\label{conjugation}
\ee
Throughout this appendix, all conjugates of fields and operators in
projective superspace are defined in this sense.

Projective superfields living in this projective superspace obey the
constraint
\be
\na_{\a} \U = 0 = \bar{\na}_{\da} \U
\label{eta-constraint} \ ,
\ee
and the restricted measure for integrating Lagrangians on
this subspace can be constructed from any differential operators linearly
independent of $\na$ and $\bar \na$. A convenient choice is the usual $N=1$
measure
\be
S=  \oint_C {{\rm d} \xi \over 2 \pi i \xi}
\; {\rm d}^4 x \; D^2 \bar{D}^2  G(\U, \bar{\U}, \xi)\ ,
\label{eq-action}
\ee
where $C$ is a contour in the $\xi$-plane that generically depends on $G$.
The constraints (\ref{eta-constraint}) guarantee that $S$ is $N=2$
supersymmetric; they can be rewritten as
\be
            D_\a \U = - \xi Q_\a \U\  , \qquad  \bar{Q}_{\da} \U = \xi
            \bar{D}_{\da} \U\ .
\label{eta_constraint2}
\ee
Projective superfields can be classified \cite{lr} as: i) $O(k)$ multiplets,
ii) rational multiplets, iii) analytic multiplets. We focus on $O(k)$
multiplets, which are polynomials in $\xi$ with powers ranging from $0$ to
$k$, and on analytic multiplets, which are analytic in some region of the
Riemann sphere.

For even $k=2p$ we impose a reality condition with respect to the
conjugation defined above (see (\ref{conjugation})). We use $\eta$
to denote a real finite order superfield
\be
\eta^{(2p)}(\xi)\equiv\frac{1}{\xi^p}\sum_{n=1}^{2p}\,\eta^{(2p)}_n\xi^n\ ,
\qquad \eta^{(2p)}={\bar \eta}^{(2p)}\ .
\label{eta_reality}
\ee
This reality condition relates different coefficients in the
$\xi$-expansion of $\eta$
\be
            \eta_{2p-n} = (-)^{p-n} \bar{\eta}_n\ .
\ee
There are various types of analytic multiplets. The {\em polar} (arctic
and antarctic) multiplets are analytic around the north and south
poles of the Riemann sphere, respectively:
\be
\U =\sum_{n=0}^\infty \U_n \xi^n \ ,\qquad
\bar{\U} = \sum_{n=0}^\infty \bar{\U}_n (- {1 \over \xi})^n \ .
\ee
The antarctic multiplet is conjugate to the arctic.

Similarly, the real {\em tropical} multiplet is the limit $p
\rightarrow \infty$ of the real $O(2p)$ multiplet $\eta^{(2p)}$.
It is analytic away from the polar regions, and can be regarded
as a sum of a part regular at the north pole and a part regular
at the south pole:
\be
\vv(\xi) = \sum_{n=-\infty}^{+\infty} \vv_n \xi^n \ ,\qquad \vv_{-n}= (-)^n
            \bar{\vv}_n\ .
\ee

The constraints (\ref{eta_constraint2}) relate the different
$\xi$-coefficient superfields
\be
D_\a \U_{n+1}= - Q_\a \U_n \ , \qquad\bar{D}_{\da} \U_n =
\bar{Q}_{\da} \U_{n+1}\ .
\ee
For any real $O(2p)$ multiplet these constraints are compatible
with the reality condition (\ref{eta_reality}). They also determine
what type of $N=1$ superfields the $\xi$-coefficients are.

Explicitly, for the $O(2)$ multiplet,
\be
\eta^{(2)}=\frac{\bar{v}}{\xi}+x-v\,\xi\ ,\label{o2mult}
\ee
where $v$ obeys $\bar{D}_{\da} v = Q_\a v = 0$ and hence projects
to an $N=1$ chiral superfield, and $x$ is real and obeys
$\bar D^2 x = Q^2 x = 0$, and hence projects to an $N=1$ real linear
superfield. This is precisely the $N=2$ tensor multiplet. Rigidly $N=2$
superconformal actions for $O(2)$  multiplets are given
by \eqn{eq-action}, where $\eta$ has conformal
weight two and $G$ is homogeneous of first degree with no
explicit $\xi$-dependence, as discussed in section 5.

Similarly, for the $O(4)$ multiplet,
\be
\eta^{(4)}=\frac{\bar{v}}{\xi^2}+\frac{\bar{s}}{\xi}+
y-s\,\xi+v\,\xi^2\ ,\label{o4mult}
\ee
where $v$ is constrained as for the $O(2)$ case, $s$ is a complex
linear superfield obeying ${\bar D}^2 s = Q^2 s = 0$ (which has one
complex physical scalar), and $y$ projects to an auxiliary real
unconstrained $N=1$ superfield. This particular off-shell
hypermultiplet is discussed extensively in \cite{itimr}.
A superconformal action for $O(4)$ multiplets (which have conformal weight 4)
is then constructed from a homogeneous function $G$ of degree $\ft12$.

{For} the arctic multiplet, only the two lowest coefficient
superfields are constrained. The other components are complex auxiliary
superfields unconstrained in $N=1$ superspace:
\be
\U = \bar{v} + \xi \bar{s} + \xi^2 r_2 + \xi^3 r_3 + \dots \ ,
\ee
where $v$ is again constrained as above, and $s$ is a
complex linear superfield as in the $O(4)$ case. The arctic multiplet is
another off-shell hypermultiplet, in this case with an infinite number of
auxiliary fields.

Finally, for the real tropical multiplet all the $\xi$-coefficient
superfields are unconstrained in $N=1$ superspace.

In general, different multiplets are adapted to different geometries,
{\it e.g.}, $O(2)$ multiplets arise when the hyperk\"ahler manifold has
commuting triholomorphic isometries.  It is believed that polar
multiplets can be used to describe arbitrary hyperk\"ahler manifolds
\cite{lr}. The real tropical multiplet is not used to describe
sigma-models, but rather arises in the projective superspace description
of the $N=2$ Yang-Mills multiplet \cite{PYM}.

Certain terms in the projective superspace Lagrangian $G$ do not contribute
to the action \eqn{eq-action}:
\be
G_{\rm trivial} = \eta^{(2)} \Big[f_1(\Y)+\bar f_1(\bar\Y)\Big]
+f_2(\Y)+\bar f_2(\bar\Y)\ ,
\label{trivial}
\ee
where $f_i$ are holomorphic functions. After the contour integral is
evaluated, the only terms that survive are either a function of $N=1$ chiral
superfields, or a function of $N=1$ chiral superfields times an $N=1$ linear
superfield, or the complex conjugates; such terms give rise only to total
derivatives.

\subsection{Isometries}

The polar multiplet has an infinite number of $N=1$ superfields;
consequently, it is difficult to extract the \Ka\ potential except in
special circumstances. On the other hand, the space of polar
multiplets has an algebraic structure: sums and products of arctic multiplets
are again arctic, as are holomorphic functions of arctic multiplets. This
allows for a very direct realization of triholomorphic isometries of the
hyperk\"ahler geometry in projective superspace: they are simply symmetries
of the projective superspace action (\ref{eq-action}) that are holomorphic
in the arctic multiplets. As we explain below, the whole process of gauging
triholomorphic isometries and performing hyperk\"ahler quotients, when
described in terms of polar multiplets in projective superspace is
essentially the same procedure as for \Ka\ quotients described in terms of
chiral superfields in
$N=1$ superspace.

We focus on the specific case that applies to the universal
hypermultiplet, as this can be trivially generalized to the much
larger class of hyperk\"ahler metrics that are hyperk\"ahler quotients of
some (flat) vector space (toric varieties and their non-abelian
generalizations). Thus we assume that the isometry is realized linearly on
a vector space with coordinates $\Y^I$ (the general case can be treated
using the methods of \cite{HKLR}):
\be
\d\, \Y^I\ =\ i\, \l^I{}_J\, \Y^J\ ,
\ee
where the parameter $\l$ is a constant matrix in some representation of the
isometry group. The invariant projective superspace Lagrangian is given by
\be
G\ =\ \bar\Y^I\eta_{IJ}\Y^J\ .
\ee
For the universal hypermultiplet $I=1,2,3$, and $\eta$ has signature
$-++$; this is the $N=2$ projective superspace rewrite of the $N=1$
superspace Lagrangian \eqn{chi-flat} of section 7, as we prove
below. This action has a rigid U(1,2) group of holomorphic isometries. Any
subgroup of this may be gauged by introducing a real tropical gauge
multiplet $\vv$ that is analogous to the $N=1$ gauge superfield
$V$, and which transforms as
\be
({{\rm e}^\vv})'{}_I{}^J = ( {\rm e}^{i\bar\l}\, {\rm e}^\vv){}_I{}^K\,
\eta_{KL}\,( {\rm e}^{-i\l}){}^L{}_M\,\eta^{MJ}\ ,
\ee
where now $\l$ has been generalized to an arctic gauge parameter, and the
conjugate gauge parameter $\bar\l$ is antarctic, as is the
conjugate multiplet $\bar\Y$. The gauge invariant action is
\be
G_\vv\ =\ \bar\Y^I\, ( {\rm e}^\vv){}_I{}^J\,\eta_{JK}\,\Y^K\ .
\label{GIlag}
\ee
One may also add $N=2$ Fayet-Iliopoulos terms for U(1) factors in the
gauge group; these take the form $c(\xi)\vv$ where $c$ is a {\it constant}
$O(2)$ multiplet, but since these break conformal invariance, they do not
interest us here.

These gauged actions and their relation to the discussion of section 7.1
may be understood most directly by going to covariant $N=1$ superspace
components.  This is analogous to the vector representation of $N=1$
Yang-Mills theory: We split the tropical gauge multiplet factors regular
at the north and south poles:
\be
{\rm e}^\vv={\rm e}^{\vv_-}\, {\rm e}^{\vv_+}\ ,\qquad \vv_+=
\sum_{n=0}^\infty \vv_{+n}\,\xi^n\ ,\qquad \vv_-=\bar \vv_+\ .
\ee
Because $\vv$ is an analytic superfield, $\nabla {\rm e}^\vv =0$, and we may
define a gauge-covariant analytic derivative $\DD$
\be
\DD\equiv \nabla + {\rm e}^{-\vv_-}(\nabla {\rm e}^{\vv_-}) =
\nabla -(\nabla {\rm e}^{\vv_+}) {\rm e}^{-\vv_+}\ .
\ee
Comparing powers of $\xi$ for both expressions, we conclude that $\DD$ has
only a constant and a linear term (just as $\nabla$), and hence defines
the $N=2$ gauge-covariant derivative (for a more detailed explanation see
\cite{PYM}). In particular, we find the covariantly chiral gauge
field strength $\WW$ by computing
\be
\{\bar\DD_{\da}(\xi_1),\bar\DD_{\db}(\xi_2)\}\ =\ \varepsilon_{\da\db}\,
(\xi_2-\xi_1)\, \WW\ .
\label{fieldstrength}
\ee
Note that $\WW$ is $\xi$ independent.
We may also define {\em covariantly} analytic polar multiplets
\be
\hat\Y\equiv {\rm e}^{\vv_+}\Y \ , \qquad \hat{\bar\Y} = \bar\Y
{\rm e}^{\vv_-}\ .
\ee
In terms of these, the gauge-invariant Lagrangian \eqn{GIlag} is just
quadratic, the $\xi$ integral \eqn{eq-action} can be trivially evaluated,
and the auxiliary superfields can be integrated out to get the
gauge-invariant $N=1$ superspace Lagrangian
\be
L_{N=1}=\hat{\bar v}{}^I\hat v{}^J\eta_{IJ}-\hat{\bar s}{}^I
\hat s{}^J\eta_{IJ}\ ,
\ee
where $\hat v{}^I$ are $N=1$ gauge-covariantly (vector representation) chiral
superfields and $\hat s{}^I$ are modified $N=1$ gauge-covariantly complex
linear superfields
\be
\bar\Dl_{\da}\hat v^I=0\ ,\qquad
\bar\Dl^2\hat s{}^I =\hat W{}^I{}_J\hat v{}^J\ .
\ee
Here $\hat W{}^I{}_J$ is the $N=1$ covariantly chiral projection
of the $N=2$ field strength $\WW$ \eqn{fieldstrength} in the representation
that acts on $\hat v{}^J$ and $\Dl$ is the $N=1$ gauge-covariant derivative.
We can go to chiral representation and replace
$\hat v,\hat s,\hat W$ with ordinary chiral and linear superfields
$v,s,W$ by introducing the $N=1$ gauge potential $V$:
\be
({\rm e}^V)_I{}^K\,\eta_{KJ}\equiv ({\rm e}^{V_-})_I{}^K\,\eta_{KL}
({\rm e}^{V_+})^L{}_J\ , \ \ \
\hat v = {\rm e}^{V_+} v\ , \ \ \ \hat s = {\rm e}^{V_+} s\ , \ \ \
\hat W = {\rm e}^{V_+} W{\rm e}^{-V_+}\ ,
\ee
where $V_\pm$ is the $N=1$ projection of the $\xi$-independent coefficients
of $\vv_\pm$. These substitutions lead to
\bea
L_{N=1}&=&(\bar v\, {\rm e}^V)^I \eta_{IJ} v{}^J-
(\bar s\, {\rm e}^V)^I\eta_{IJ}s{}^J\ ,\ \label{lagn1s} \\ [3mm]
\bar D^2 s^I &=& W{}^I{}_J v^J\ . \label{sconst}
\eea
It is convenient to rewrite the $N=1$ Lagrangian \eqn{lagn1s} in terms of
chiral superfields; to do this, we impose the constraints \eqn{sconst} by
chiral Lagrange multipliers $z_{-I}$ in a superpotential term
\be
z_{-I}(\bar D^2 s^I - W{}^I{}_J v^J)\ ,
\ee
and integrate out $s$ to obtain the non-abelian generalization of the $N=1$
gauged Lagrangian \eqn{chi-gauged} (after relabeling $v\to z_+$):
\be
L_{N=1}= (\bar z_+\, {\rm e}^V)^I \eta_{IJ} z_+{}^J-
z_{-I} \eta^{IJ}({\rm e}^{-V}\bar z_-)_J\ . \label{Ln=1-gauged}
\ee
In addition, we are left with a superpotential
term
\be
\Tr \,[\,W \mu^+]=z_{-I}W^I{}_Jz_{+}^J\ ,
\ee
where $\mu^+$ is just (the non-abelian generalization of) the holomorphic
moment map \eqn{moment-U1}. Observe that interchanging $z_+\leftrightarrow
z_-$ and changing the representation of $V$ to its conjugate does not modify
the gauged Lagrangian \eqn{Ln=1-gauged}; this implies that in the original
$N=2$ Lagrangian $G_\vv$ \eqn{GIlag}, we can take $\Y$ transforming in the
conjugate representation ({\it e.g.}, opposite charge for U(1)) without
changing the final result. In the next subsection we integrate out the $N=2$
gauge fields to find the quotient Lagrangian; in $N=1$ superspace,
integrating out the chiral superfield $W$ imposes the moment map constraint
$\mu^+=0$.

\subsection{Quotients and Duality}

Just as $N=2$ isometries and gauging in projective superspace bear a
striking resemblance to their $N=1$ superspace analogs, so do $N=2$
quotients and duality; indeed, the tensor multiplet projective superspace
Lagrangian is just the Legendre transform of the polar multiplet
Lagrangian.

The procedure we follow is the same as in $N=1$ superspace: we gauge
the relevant isometries as above; to perform a quotient, we simply
integrate out the gauge prepotential ${\rm e}^\vv$. Since this
does not break the isometry, we are left with an action defined on the
quotient space. To find the dual, we add a Lagrange multiplier $\eta$ that
constrains the gauge prepotential to be trivial\footnote{As explained in
\cite{hklr2,rv}, this is the correct geometric way of understanding duality;
when one chooses coordinates such that the Killing vectors generating the
isometries are constant, this gives the usual Legendre transform.}, and
again integrate out $\vv$; the dual field is then the Lagrange multiplier
$\eta$. As in the $N=1$ case, we only consider duality for abelian
isometries. In that case, the Lagrange multiplier term that constrains $\vv$
is
\be
\eta \vv\ ,
\ee
where $\eta$ is the $O(2)$ superfield that describes the $N=2$ tensor
multiplet as explained above \eqn{o2mult}.

We have now assembled all the tools we need to find the various tensor
multiplet formulations of the universal hypermultiplet.

The HKC of the universal hypermultiplet is just the U(1) hyperk\"ahler
quotient of a flat Lorentzian space:
\be
\hat G_\vv=(-\bar\Y^1\Y^1+\bar\Y^2\Y^2+\bar\Y^3\Y^3)\,{\rm e}^\vv\ .
\label{GU12}
\ee
This is manifestly invariant under U(1,2); however, if we
integrate out $\vv$, we get a constraint (the moment map
constraints in projective superspace) rather than an $N=2$ superspace
Lagrangian -- even though we do get a corresponding $N=1$ chiral superfield
quotient Lagrangian from \eqn{Ln=1-gauged}.  We can get an $N=2$ quotient
Lagrangian at the expense of losing manifest U(1,2) invariance by using
the observation from the end of the previous subsection that the final $N=1$
superspace result does not change when we take $\vv\to -\vv$:
\be
\check G_\vv=(-\bar\Y^1\Y^1+\bar\Y^2\Y^2)\,{\rm e}^\vv
+\bar\Y^3\Y^3\,{\rm e}^{-\vv}\ .
\label{GU11}
\ee
Integrating out $\vv$ and fixing the gauge $\Y^3=1$ gives the polar
multiplet superspace Lagrangian for the hyperk\"ahler cone of the universal
hypermultiplet:
\be
G_{\rm HKC}=2\sqrt{-\bar\Y^1\Y^1+\bar\Y^2\Y^2}\ ,
\label{GHKCUH}
\ee
which has manifest U(1,1) rather than SU(1,2) invariance.

We now dualize \eqn{GU12} with respect to any pair of commuting holomorphic
isometries as discussed above by gauging the isometries and adding a
term $\eta^1\vv_1+\eta^2\vv_2$. The three choices of inequivalent
commuting pairs of U(1) generators are given in \eqn{gen1}-\eqn{gen3}.
Exponentiating the generators gives three different expressions:
\bea
{\rm e}^{\vv_1T_1+\vv_2T_2} &=& \pmatrix {{\rm e}^{-\vv_1-\vv_2} & 0 & 0 \cr
0 & {\rm e}^{-\vv_1+\vv_2} & 0 \cr 0 & 0 & {\rm e}^{2\vv_1}}\ ,
\label{vv1}\\[5mm]
{\rm e}^{\vv_1T_1+\vv_2T'_2} &=& \pmatrix {{\rm e}^{-\vv_1} & 0 & 0 \cr 0 &
{\rm e}^{-\vv_1} & 0 \cr 0 & 0 & {\rm e}^{2\vv_1}}
\pmatrix{1+\vv_2 & \vv_2 & 0
\cr -\vv_2 & 1-\vv_2 & 0 \cr 0 &0 & 1}\ ,
\label{vv2}\\[5mm]
{\rm e}^{\vv_1T'_1+\vv_2T'_2} &=&
\pmatrix {1-\ft12 \vv_1^2 & -\ft12 \vv_1^2 &
-\vv_1 \cr \ft12 \vv_1^2 & 1+\ft12 \vv_1^2 & \vv_1
\cr \vv_1 &  \vv_1 & 1}
\pmatrix{1+\vv_2 & \vv_2 & 0
\cr -\vv_2 & 1-\vv_2 & 0 \cr 0 &0 & 1}\nonumber\\[3mm]
&=&\pmatrix {1+\vv_2-\ft12 \vv_1^2 &\vv_2 -\ft12 \vv_1^2 &
-\vv_1 \cr-\vv_2+ \ft12 \vv_1^2 & 1-\vv_2+\ft12 \vv_1^2 & \vv_1
\cr \vv_1 &  \vv_1 & 1}\ .
\label{vv3}
\eea
It is now straightforward to evaluate the gauged action $\hat G_\vv$
\eqn{GU12}, add $\eta^1\vv_1+\eta^2\vv_2$, and integrate out $\vv$, $\vv_1$
and $\vv_2$.  Up to signs that can be fixed by choosing the orientations of
the contours, trivial terms \eqn{trivial}, and linear redefinitions of
the $\eta^I$, we find the three dual Lagrangians \eqn{G1}, \eqn{G2}, and
\eqn{G3}.

One may also dualize the U(1,1) invariant Lagrangian $G_{\rm HKC}$
\eqn{GHKCUH}. Here there are only two choices of inequivalent pairs of
commuting generators: the $2\times 2$ identity matrix ${\bf 1}_2$ and one of
\be
\pmatrix{1 & 0 \cr 0 & -1}\qquad {\rm or}\qquad
\pmatrix{1 & 1 \cr -1 & -1}\ .
\ee
Again up to signs having to do with the orientations of the contours, as
well as irrelevant terms, these give only the two Lagrangians \eqn{G1} and
\eqn{G2}.

It is also instructive to consider the example of anti-de-Sitter space
$AdS_4$.  The HKC is eight-dimensional flat space with a polar multiplet
Lagrangian
\be
G_{AdS_4} = -\bar\Y^1\Y^1+\bar\Y^2\Y^2\ ,
\ee
that is U(1,1) invariant. This can be dualized with respect to the
same inequivalent pairs of commuting generators as in the previous
paragraph; the resulting tensor actions are
\be
G_{AdS_4}^1 = -\eta^1\ln\eta^1 + \eta^2\ln\eta^2\ ,\qquad
G_{AdS_4}^2 = -\eta^1\ln\eta^2\ ,
\ee
respectively.

\eject
%

\end{document}